\documentclass[12pt, draftclsnofoot,onecolumn, letterpaper]{IEEEtran}
\usepackage[T1]{fontenc}% optional T1 font encoding
\usepackage{color}
\usepackage{amsmath}
\usepackage{capt-of}
\usepackage{url}
\usepackage{array}
\usepackage{fancyhdr}
\usepackage{latexsym}
\usepackage{diagbox}
\usepackage{epsfig}
\usepackage{amssymb}
\usepackage{amsfonts}
\usepackage{amsxtra}
\usepackage{xspace}
\usepackage{makeidx}
\usepackage{graphics}
\usepackage{pdfpages}
\usepackage{theorem}
\usepackage{makecell,rotating,multirow,diagbox}
\usepackage{subfigure}
\usepackage{algorithm}
\usepackage{algorithmic}
\usepackage{multirow}
\usepackage[]{hyperref}
\usepackage{cite}
\DeclareMathOperator*{\minimize}{minimize}

\newtheorem{lemma}{\textbf{Lemma}}

\begin{document}
\title{Channel Estimation for IRS-Assisted Millimeter-Wave  MIMO Systems: Sparsity-Inspired Approaches}

\author{Tian~Lin,~\IEEEmembership{Student Member,~IEEE,} Xianghao~Yu,\IEEEmembership{ Member,~IEEE,} Yu~Zhu,~\IEEEmembership{Member,~IEEE,} and Robert~Schober,~\IEEEmembership{Fellow,~IEEE}
\thanks{This paper was presented in part at the IEEE Global Communications Conference,
Taipei, Taiwan, Dec. 2020 \cite{lin_channel_2020}.}
\thanks{Tian Lin and Yu Zhu are with the Department of Communication Science and Engineering, Fudan University, Shanghai, China (e-mail: lint17@fudan.edu.cn, zhuyu@fudan.edu.cn).}
\thanks{Xiagnhao Yu is with the Department of Electronic and Computer Engineering, the Hong Kong University of Science and Technology (HKUST), Kowloon, Hong Kong (e-mail: eexyu@ust.hk).}
\thanks{Robert Schober is with the Institute for Digital Communications, Friedrich-Alexander-University Erlangen-Nuremberg (FAU), 91054 Erlangen, Germany (e-mail: robert.schober@fau.de).}}

\maketitle

%=======================================
%               Abstract
%=======================================
\setlength{\abovecaptionskip}{-4pt}
\begin{abstract}
Due to their ability to create favorable line-of-sight (LoS) propagation environments, intelligent reflecting surfaces (IRSs) are regarded as promising enablers for future millimeter-wave (mm-wave) wireless communication. In this paper, we investigate  channel estimation for IRS-assisted mm-wave multiple-input multiple-output (MIMO) {\color{black}wireles}s systems. By leveraging the sparsity of mm-wave channels in the angular domain, we formulate the channel estimation problem as an $\ell_1$-norm regularized  optimization problem with fixed-rank constraints.  To tackle the non-convexity of the formulated problem, an efficient algorithm is proposed by capitalizing on alternating minimization and manifold optimization (MO),  which  yields a locally optimal solution. To further reduce the computational complexity of the estimation algorithm, we propose a compressive sensing- (CS-)  based channel estimation approach. In particular, a three-stage estimation protocol is put forward where the subproblem in each stage can be solved via  low-complexity CS methods. Furthermore, based on the acquired channel state information (CSI)  of the cascaded channel, we design a passive beamforming algorithm for  maximization of the  spectral efficiency. Simulation results reveal that the proposed MO-based estimation (MO-EST)  and beamforming algorithms significantly outperform two benchmark schemes while the CS-based estimation (CS-EST)  algorithm strikes a balance between performance and complexity. In addition, we demonstrate the robustness of the MO-EST algorithm with respect to imperfect knowledge of the sparsity level of the channels, which is crucial for practical implementations.

\end{abstract}

\IEEEpeerreviewmaketitle

%=======================================
%               Introduction
%=======================================

\section{Introduction}\label{sec:introduction}
Due to its enormous potential for overcoming the spectrum crunch, millimeter-wave (mm-wave) communication has been regarded as a key technology for future wireless cellular systems \cite{Pi2011Intro, Wang2015mmwave}. By leveraging large antenna arrays to synthesize directional beams  and exploiting the large available bandwidth, mm-wave communication enables gigabit-per-second data rates \cite{Rangan2014mmwave}. However, mm-wave communication is vulnerable to blockages due to the  limited scattering effects at mm-wave frequencies.  As the propagation environment of conventional mm-wave communication systems is uncontrollable, the quality of service (QoS) is significantly degraded when  line-of-sight (LoS) links are not available. \par

Recently, intelligent reflecting surfaces (IRSs) have been incorporated into  wireless communication systems, mainly due to their capability of customizing favorable wireless propagation environments \cite{ wu2020towards, huang2019RIS}. Equipped with a large number of low-cost \textit{passive} reflecting elements, e.g., diodes and phase shifters,  IRSs enable the adaptation of wireless propagation environments  with limited power consumption \cite{wu2020intelligent}. This property of IRSs is particularly advantageous for  coverage extension of mm-wave wireless systems  \cite{Nemati2020, Pradhan2020, Yu2021}. Specifically,  when the direct LoS links between the base station (BS) and the user equipments (UEs)  are blocked, the deployed IRSs can reflect the incident signals to provide virtual LoS links for mm-wave communications. Furthermore, with  well-designed IRS reflecting elements, the communication performance can be further enhanced via programmable and reconfigurable signal reﬂections. Inspite of {\color{black}their} great potential,  the introduction of IRSs in wireless systems also brings  new challenges,  among which the acquisition of channel state information (CSI)  may be the most difficult task.  Although the CSI of the direct BS-UE links can be  obtained by turning off the IRSs  and applying conventional CSI acquisition approaches, it is difficult to estimate the two  IRS-assisted channels, i.e., the BS-IRS channel and  IRS-UE channel. In particular, since radio frequency (RF) chains are not available at the passive IRSs, it is not possible to estimate the two IRS-assisted channels directly by regarding the IRS as a conventional RF chain-driven transceiver. Therefore, the classical channel estimation techniques are not applicable to the {\color{black}newly-emerging} IRS-assisted communication systems.\par

Several works have investigated channel estimation in IRS-assisted wireless systems \cite{wang_channel_2020,jensen_optimal_2019, chen2019channel, taha2019enabling, liu2020Matrix, He2020Cascaded, he2020channel, araujo_parafac-based_2020, Wei2020ParaFAC,  wang_compressed_2020,  ma_joint_2020}.  The authors of \cite{wang_channel_2020, jensen_optimal_2019} characterized the minimum pilot sequence length for channel estimation in IRS-assisted multi-user multiple-input single-output (MISO) systems based on the least square (LS) criterion. To further reduce the pilot overhead,   compressive sensing (CS) techniques were utilized in \cite{chen2019channel, taha2019enabling, liu2020Matrix}  to solve the estimation problem  based on the assumption that the channel matrices are sparse. However, the  algorithms proposed in these works are only applicable in  wireless systems with single-antenna UEs, which are unlikely to be used in mm-wave systems. Channel estimation {\color{black}for} IRS-assisted multiple-input multiple-output (MIMO) systems was first studied in \cite{He2020Cascaded}. A matrix factorization-based algorithm was proposed where {\color{black}during training} each reflecting element was turned on successively while keeping the remaining reflecting elements off. In \cite{he2020channel}, the authors focused on the estimation of the dominant LoS path of  IRS-assisted MIMO channels to simplify the channel estimation problem, and an iterative reweighting method based on the gradient descend algorithm was proposed.  By modeling the received pilots of IRS-assisted MIMO system as a tensor, a parallel factor decomposition (PARAFAC) algorithm was developed in \cite{araujo_parafac-based_2020, Wei2020ParaFAC}  to estimate the IRS-assisted channels. While these approaches designed for sub-6 GHz bands are also applicable to mm-wave MIMO systems,  a significant performance loss is expected as the unique channel characteristics of mm-wave MIMO systems are not taken into account.   By exploiting the sparsity of  mm-wave channels,  the channel estimation {\color{black}for} IRS-assisted MIMO mm-wave systems was formulated as a classical sparse signal recovery problem in \cite{wang_compressed_2020}, such that conventional CS techniques, e.g., the orthogonal matching pursuit (OMP) and the generalized approximate message passing (GAMP) algorithms,  can be directly applied. In addition, an iterative atom pruning based subspace pursuit (IAP-SP) scheme was developed  to solve the sparse signal recovery problem, where the columns of the sensing matrix that are least correlated with the signal residual are eliminated in the iterative process \cite{ma_joint_2020}. However, the computational complexity of these CS-based algorithms  scales cubically in  the {\color{black}number of antennas and the number} of reflecting elements \cite{wang_compressed_2020, ma_joint_2020}, which is prohibitively high especially in massive MIMO systems empowered by large-scale IRSs. Therefore, {\color{black}we conclude that} efficient channel estimation algorithms for IRS-assisted mm-wave MIMO systems are not available in the literature, yet.

%The required training overhead is associated with the sampling rate of the observation matrix, i.e.,  we can strike a good balance between the estimation performance and training overhead by selecting an appropriate  sampling rate.
In this paper, we investigate the  channel estimation problem of  IRS-assisted mm-wave MIMO systems operating in the time division duplex (TDD) mode. The BS and UE are both equipped with multiple antennas, and an IRS consisting of programmable phase shifters is deployed to customize a favorable propagation environment for mm-wave communication. {\color{black}Compared to its conference version \cite{lin_channel_2020}, this paper leverages the sparsity of   mm-wave channels in the angular domain}, based on which the channel estimation problem can be  formulated as an $\ell_1$-norm regularized optimization problem with fixed-rank constraints. To tackle the high degree of non-convexity  of the formulated problem, we first apply  alternating minimization (AM)  to decouple the formulated problem into two subproblems,  which correspond to the estimation of the UE-IRS channel and the IRS-BS channel, respectively.  Subsequently, manifold optimization (MO)  is employed to address the fixed-rank constraint in each one of the subproblems, which leads to  a locally optimal solution  of the channel estimation problem. To further reduce the computational complexity, we  propose a  channel estimation algorithm based on CS techniques. In particular,  we divide the overall estimation phase into three stages, where the subproblem in each stage can be efficiently solved by the OMP method. Finally, with the estimated channel at hand, a novel passive beamforming algorithm for  spectral efficiency maximization is developed by solving an equivalent weighted  mean square  error  minimization (WMMSE) problem. Based on the AM principle, closed-form solutions of the beamformers at the BS and UE are derived while the reflection coefficients are optimized via the MO technique. The proposed beamforming algorithm is guaranteed to converge to a locally optimal solution of the WMMSE problem.  Simulation results  show that the proposed MO-based estimation (MO-EST) algorithm significantly outperforms two benchmark schemes  proposed in  \cite{araujo_parafac-based_2020} and \cite{wang_compressed_2020}, respectively. Meanwhile, the CS-based estimation (CS-EST) algorithm strikes a good balance between  performance and computational complexity. Besides, we  demonstrate the robustness of the proposed estimation algorithms when the channel sparsity levels are not perfectly known. We also reveal the superiority of the proposed beamforming algorithm compared with the state of the art \cite{Pan2020Rate, Wang2021Rate}.

\emph{Notations:} In this paper, the imaginary unit of a complex
number is denoted by $\jmath=\sqrt{-1}$.
The set of nonnegative integers is denoted by $\mathbb{N}=\{0,1,\ldots\}$. $\mathbb{C}^{m\times n}$ denotes the set of all $m \times n$ complex-valued matrices. Matrices and vectors are denoted by boldface capital and lower-case letters, respectively. $\mathbf{I}_N$ denotes the $N \times N$ identity matrix. $\mathbf{1}_{N}$ denotes the $N\times 1$ all-ones vector. $(\cdot)^*$, $(\cdot)^T$, $(\cdot)^H$, $\mathrm{rank}(\cdot)$, $\mathrm{tr(\cdot)}$, $\mathrm{vec(\cdot)}$, and $\|\cdot\|_F$ denote the conjugate, transpose, conjugate transpose, rank, trace, vectorization, and Frobenius norm of a matrix, respectively. $\|\cdot\|_0$, $\|\cdot\|_1$, and $\|\cdot\|$ represent the $\ell_0$-norm, $\ell_1$-norm, and $l_2$-norm of a vector, respectively.
The Hadamard, Kronecker, and Khatri-Rao products are represented by  $\circ$, $\otimes$, and $\odot$, respectively.  $|\cdot|$ denotes the absolute value or the magnitude of a complex number. {\color{black}$\mathrm{d}(\cdot$) denotes the differential, i.e., an  infinitesimal difference in calculus}.
$\Re(\cdot)$ and $\mathbb{E}(\cdot)$ denote the real part of a complex number and statistical expectation, respectively.    $\mathrm{diag}(\mathbf{x})$  is a diagonal matrix with the entries of $\mathbf{x}$ on its main diagonal. $\mathcal{CN}(\mathbf{0}, \mathbf{\Sigma})$ denotes  the  circularly symmetric complex Gaussian distribution with zero mean and  covariance matrix $\mathbf{\Sigma}$. $[\mathbf{A}]_{ij}$ and $[\mathbf{a}]_{i}$ denote the $(i,j)$-th entry of  matrix $\mathbf{A}$ and the $i$-th entry of  vector $\mathbf{a}$, respectively.

%=======================================
%               Sec2: System Model
%=======================================

\begin{figure}[!t]
 		\centering
 		\includegraphics[height=2.0in]{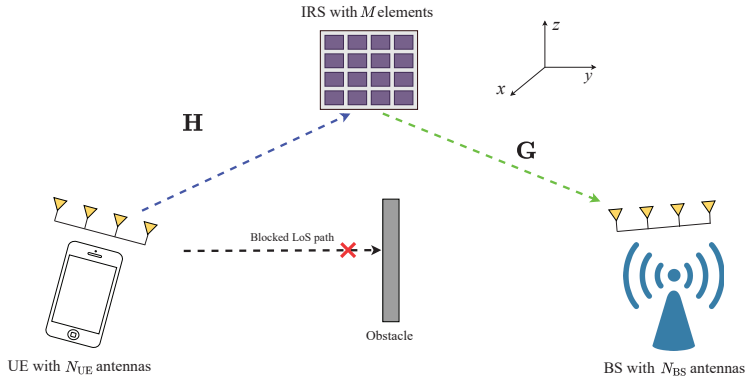}
 	\caption{ An uplink IRS-assisted mm-wave MIMO communication system.}
 	    \label{fig:system}
 	\end{figure}
\section{System Model and Channel Model}\label{sec:System-model}
\subsection{System Model}\label{subsec:Channel-estimation-protocol}
We consider  uplink channel estimation in an IRS-assisted mm-wave  point-to-point MIMO system operating in the TDD mode, as shown in Fig. \ref{fig:system}.  The UE and the BS are equipped with $N_\mathrm{UE}$ and $N_\mathrm{BS}$ antennas, respectively. In addition, one passive IRS employing $M$ phase shifters is deployed to establish a virtual LoS link for the UE that would otherwise be blocked\footnote{The proposed scheme can be readily extended to the scenario with LoS links. In particular, we can switch off all the IRS elements \cite{wang_channel_2020}, then the estimation of the direct UE-BS channel reduces to a conventional MIMO channel estimation problem, which can be efficiently solved via existing algorithms \cite{JLee2016}.}. For each block, we assume that the time period available for channel estimation is divided into $T$ time slots. In the $t$-th time slot,  the equivalent baseband received signal $\mathbf{r}_t\in\mathbb{C}^{N_\mathrm{BS}}$ at the BS side can be expressed as {\color{black}follows}
\begin{equation}\label{eqn:rt}
    \mathbf{r}_t = \mathbf{G}\mathbf{\Phi}_t\mathbf{H}\mathbf{s}_t + \mathbf{z}_t,
\end{equation}
where $\mathbf{s}_t\in\mathbb{C}^{N_\mathrm{UE}}$ denotes the transmit pilot vector in the $t$-th time slot, which is  known at the BS, and the power allocated to the pilot vector is given by $\|\mathbf{s}_t\|^2=P_\mathrm{tr}$. 
$\mathbf{H}\in\mathbb{C}^{M\times N_\mathrm{UE}}$ and $\mathbf{G}\in\mathbb{C}^{N_\mathrm{BS}\times M}$ represent the channels from the UE to the IRS and from the IRS to the BS, respectively. $\mathbf{z}_t$ denotes the received additive Gaussian noise vector with  $\mathbf{z}_t \sim \mathcal{CN}(\mathbf{0}, \sigma^2\mathbf{I}_{N_\mathrm{BS}})$, where $\sigma^2$ represents the noise power. The reflection coefficient matrix of the IRS in the $t$-th time slot is given by  diagonal matrix $\mathbf{\Phi}_t = \mathrm{diag}(\mathbf{v}_t)$, where $\mathbf{v}_t\in \mathbb{C}^{M}$ represents the training reflection coefficient vector. Since the IRS is implemented by phase shifters \cite{yu_enabling_2019},  the reflecting elements can only change the phases of the signals, i.e., $|[\mathbf{v}_t]_n|=1$ for $n=1,\ldots, M$. Finally, the  pilots received in all $T$ time slots, i.e., $\mathbf{r}_1, \ldots, \mathbf{r}_{T}$, are collected at the BS for channel estimation.
 
 \subsection{Mm-Wave Channel Model}\label{sec:mmwave-model}
 In this paper, we assume that the UE and the BS are both equipped with   uniform  linear array (ULA) antennas. As the mm-wave propagation environment is  well characterized by the Saleh-Valenzuela model \cite{wang_compressed_2020,chen2019channel}, the channel matrices can be modeled as
\begin{equation}
\label{channel_model}
\begin{split}
 &\mathbf{G} =\sqrt {\frac{{{N_{\rm{BS}}}{M}}}{P}} \sum\limits_{p = 1}^{P} {{\alpha _{p}}{\mathbf{a}_{{\mathrm{BS}}}}(\theta _{\mathrm{r}}^{p}){\mathbf{a}_\mathrm{IRS}^H}(\theta _{\mathrm{t}}^{p},\phi _{\mathrm{t}}^{p}} {)}, \\
 &\mathbf{H} =\sqrt {\frac{{{N_{\rm{UE}}}{M}}}{Q}} \sum\limits_{q = 1}^{Q} {{\beta _{q}}{\mathbf{a}_{\rm{IRS}}}(\psi _{\mathrm{r}}^{q},\varphi _{\mathrm{r}}^{q}){\mathbf{a}_{\mathrm{UE}}^H}(\psi _\mathrm{t}^{q})},
 \end{split}
\end{equation}
where $\alpha_{p}$, $\theta_\mathrm{r}^{p}$,  and $\theta_\mathrm{t}^{p}$ ($\phi_\mathrm{t}^{p}$) denote the complex gain, angle of arrival (AoA), and azimuth (elevation) angle of departure (AoD) of the $p$-th path of the IRS-BS channel. Similarly, $\beta_{q}$, $\psi_\mathrm{r}^{q}$ ($\varphi_\mathrm{r}^{q}$), and $\psi_\mathrm{t}^{q}$ denote  the complex gain, azimuth (elevation) AoA, and  AoD of the $q$-th path of the UE-IRS channel. In addition, $P$ and $Q$ denote the numbers of paths of the IRS-BS channel and the UE-IRS channel, respectively. Besides, $\mathbf{a}_\mathrm{BS}$, $\mathbf{a}_\mathrm{IRS}$, and $\mathbf{a}_\mathrm{UE}$ denote the receive and transmit array response vectors at the BS, IRS, and UE,  respectively. Specifically, define 
\begin{equation}
    \mathbf{f}(u, N) = \frac{1}{\sqrt{N}}[1, e^{\jmath\pi u}, \dots, e^{\jmath\pi(N-1)u}]^T.
\end{equation}
Then, the array response vectors of the half-wavelength spaced ULAs at the BS and UE are  given by
\begin{equation}\label{ULA}
    \mathbf{a}_\mathrm{BS}(\theta_\mathrm{r}^p) = \mathbf{f}\left(\cos\left(\theta_\mathrm{r}^p\right), N_\mathrm{BS}\right) \quad \mathrm{and} \quad \mathbf{a}_\mathrm{UE}(\psi _\mathrm{t}^{q})=\mathbf{f}\left(\cos\left(\psi _\mathrm{t}^{q}\right), N_\mathrm{UE}\right),
\end{equation}
respectively. In addition, the array response vector of the IRS involving $M_z\times M_y$ ($M\triangleq M_yM_z$) elements is given by
\begin{equation}
\label{USPA}
    \mathbf{a}_\mathrm{IRS}(\theta, \phi) = \mathbf{a}_y(\theta, \phi) \otimes \mathbf{a}_z(\phi),
\end{equation}
where $\mathbf{a}_y(\theta, \phi) = \mathbf{f}(\sin\theta\sin\phi, M_y), \quad\mathbf{a}_z(\phi) =\mathbf{f}(\cos\phi, M_z).$

\emph{Remark 1:} In {\color{black}Sections III and  IV}, the numbers of paths $P$ and $Q$ are  assumed to be  known  at the BS. In practice, they can be estimated by classical direction-of-arrival (DOA) estimation methods,  e.g., multiple signal classification (MUSIC) and estimation of signal parameters via rotational invariant techniques (ESPRIT). In Section VI, we shall consider the case where $P$ and $Q$ are not accurately known  to test the robustness of the proposed algorithms with respect to a mismatched number of paths.

\subsection{Sparse Representation  of Mm-Wave Channel}\label{sparse}
%  by ignoring the angle quantization error \cite{Heath2016},
Considering the described channel model,  the two IRS-assisted mm-wave channel matrices to be estimated can be  rewritten in an angular domain representation as follows \cite{chen2019channel, wang_compressed_2020}  
\begin{equation}\label{eqn:VAD}
             \mathbf{H} = \mathbf{A}_\mathrm{I}\mathbf{\Lambda}_\mathbf{H}\mathbf{A}_\mathrm{UE}^H,\quad
            \mathbf{G} = \mathbf{A}_\mathrm{BS}\mathbf{\Lambda}_\mathbf{G}{\mathbf{A}}_\mathrm{I}^H,  
\end{equation}
where $\mathbf{A}_\mathrm{I}\in \mathbb{C}^{M\times G_\mathrm{I}}$, $\mathbf{A}_\mathrm{UE}\in \mathbb{C}^{N_\mathrm{UE}\times G_\mathrm{UE}}$, and $\mathbf{A}_\mathrm{BS} \in \mathbb{C}^{N_\mathrm{BS}\times G_\mathrm{BS}}$ are three overcomplete dictionary matrices in the angular domain consisting of array response vectors, each of which corresponds to one specific AoA/AoD at the {\color{black}IRS, UE, and BS}, respectively \cite{chen2019channel}. Here, $G_\mathrm{I}$, $G_\mathrm{UE}$, and $G_\mathrm{BS}$  represent the corresponding angular resolutions. Thus, ${\mathbf{\Lambda}}_\mathbf{G}\in \mathbb{C}^{{G}_\mathrm{BS}\times {G}_\mathrm{I}}$ and $\mathbf{\Lambda}_\mathbf{H}\in \mathbb{C}^{G_\mathrm{I}\times G_\mathrm{UE}}$ are two  angular domain sparse matrices with $P$ and $Q$ non-zero elements corresponding to the channel path gains $\{\alpha_p\}$ and $\{\beta_q\}$ in \eqref{channel_model}, respectively \cite{XLi2018, ma_joint_2020}.  Specifically, according to \eqref{ULA}, the dictionary matrices $\mathbf{A}_\mathrm{BS}$ and $\mathbf{A}_\mathrm{UE}$ are given by 
\begin{equation}\label{BSUEcode}
    \begin{split}
        \mathbf{A}_\mathrm{BS} &= \left[\mathbf{f}(-1, N_\mathrm{BS}), \mathbf{f}(-1 + \frac{2}{G_\mathrm{BS}}, N_\mathrm{BS}), \ldots, \mathbf{f}(1 - \frac{2}{G_\mathrm{BS}}, N_\mathrm{BS})\right],\\
        \mathbf{A}_\mathrm{UE} &= \left[\mathbf{f}(-1, N_\mathrm{UE}), \mathbf{f}(-1 + \frac{2}{G_\mathrm{UE}}, N_\mathrm{UE}), \ldots, \mathbf{f}(1 - \frac{2}{G_\mathrm{UE}}, N_\mathrm{UE})\right].
    \end{split}
\end{equation}Similarly, according to \eqref{USPA}, the  dictionary matrix $\mathbf{A}_\mathrm{I}$ is given by
\begin{equation}\label{eqn:UPA_dictionary}
    \mathbf{A}_\mathrm{I} = \mathbf{A}_y\otimes\mathbf{A}_z,
\end{equation}
where $\mathbf{A}_y = [\mathbf{f}(-1, M_y), \mathbf{f}(-1 + \frac{2}{G_y}, M_y), \ldots, \mathbf{f}(1 - \frac{2}{G_y}, M_y)]$ and $\mathbf{A}_z = [\mathbf{f}(-1, M_z), \mathbf{f}(-1 + \frac{2}{G_z}, M_z), \ldots, \mathbf{f}(1 - \frac{2}{G_z}, M_z)]$. Here, $G_y$ and $G_z$ denote the angular resolutions along the $y$- and $z$-axes, respectively, such that $G_\mathrm{I} = G_yG_z$. 
% \subsection{Estimation problem formulation} \label{sec:HBF-design}

%=======================================
%         MO-algorithm
%=======================================

\section{Proposed MO-EST Algorithm}
In this section,  we first exploit the sparsity of mm-wave channels and formulate the estimation problem. Then, by capitalizing on the AM and MO techniques, we develop the MO-EST algorithm to efficiently obtain a locally optimal solution  for the formulated estimation problem.
\subsection{Estimation Problem Formulation} \label{sec:HBF-design}
According to \cite{jensen_optimal_2019}, the minimum variance unbiased estimators of $\mathbf{H}$ and $\mathbf{G}$  can be obtained based on the LS criterion. Thus, based on the system model elaborated in Section II, the LS estimators of the channels can be obtained by solving the following problem
\begin{equation}\label{eqn:ls-func}
\begin{array}{cl}
\displaystyle{\minimize_{\hat{\mathbf{G}}, \hat{\mathbf{H}}}} &  \sum_{t=1}^{T}{\|\mathbf{r}_t - \hat{\mathbf{G}}\mathbf{\Phi}_t\hat{\mathbf{H}}\mathbf{s}_t\|^2},
\end{array}
\end{equation}
where $\hat{\mathbf{G}}$ and $\hat{\mathbf{H}}$ denote the estimates of $\mathbf{G}$ and $\mathbf{H}$, respectively. Note that this LS formulation holds for any IRS-assisted wireless system. However, directly solving  problem \eqref{eqn:ls-func} does not leverage the special properties of mm-wave channels, which inevitably leads to  a significant performance loss \cite{lin_channel_2020}. Besides,  $T\ge MN_\mathrm{UE}$ is required to guarantee a unique solution of problem \eqref{eqn:ls-func} \cite{jensen_optimal_2019, araujo_parafac-based_2020}. In other words,  the training overhead becomes prohibitive for  large numbers of  antennas and  reflecting elements.   Therefore, before formulating the channel estimation problem, we explicitly exploit two unique properties of mm-wave channels in the following two lemmas.
\begin{lemma} \label{lemma1}
 Suppose $\mathrm{min}(N_\mathrm{BS},  M) \ge P$ and $\mathrm{min}(N_\mathrm{UE},  M) \ge Q$, then we have
\begin{equation}
    \mathrm{rank}(\mathbf{G}) = P, \quad \mathrm{rank}(\mathbf{H}) = Q.
\end{equation}
\end{lemma}
\textit{Proof}: Please refer to Appendix A. $\hfill\blacksquare$

Lemma 1 indicates {\color{black}that}  mm-wave channels have a fixed low rank, i.e., the channel matrices are sparse in their eigenvalues. It was shown in \cite{lin_channel_2020} that  mm-wave channels can be effectively estimated by leveraging their low-rank property. However, the consideration of the low-rank constraint in estimation problem \eqref{eqn:ls-func} does not reduce the exceeding{\color{black}ly} large training overhead, i.e., {\color{black}we still need} $T\ge MN_\mathrm{UE}$. Therefore, we {\color{black}also exploit} another crucial property of mm-wave channels. Based on the angular domain representation of $\mathbf{H}$ and $\mathbf{G}$ in Section \ref{sparse},  we present the following lemma.
\begin{lemma} \label{lemma1}
 By setting   $G_\mathrm{I}= M$, $G_\mathrm{UE}= N_\mathrm{UE}$, and $G_\mathrm{BS}= N_\mathrm{BS}$, the two $\ell_0$-norms related to the estimated channels are asymptotically given by
\begin{equation}\label{lemma2-eqn}
     \|\boldsymbol{\lambda}_\mathbf{H}\|_0=Q,\quad \|\boldsymbol{\lambda}_\mathbf{G}\|_0=P,
\end{equation}
where $\boldsymbol{\lambda}_\mathbf{H}\triangleq\mathrm{vec}\left(\mathbf{A}_\mathrm{I}^H\mathbf{H}\mathbf{A}_\mathrm{UE}\right)$ and $\boldsymbol{\lambda}_\mathbf{G}\triangleq\mathrm{vec}\left(\mathbf{A}_\mathrm{BS}^H\mathbf{G}\mathbf{A}_\mathrm{I}\right).$
\end{lemma}
\textit{Proof}: Notice that when $G_\mathrm{I}= M$, $G_\mathrm{UE}= N_\mathrm{UE}$, and $G_\mathrm{BS}= N_\mathrm{BS}$,  $\mathbf{A}_\mathrm{BS}$, $\mathbf{A}_\mathrm{UE}$, and $\mathbf{A}_\mathrm{I}$ are all unitary matrices according to \eqref{BSUEcode} and \eqref{eqn:UPA_dictionary}. Therefore, we have $\mathbf{A}_\mathrm{I}^H\mathbf{H}\mathbf{A}_\mathrm{UE}=\mathbf{\Lambda}_\mathbf{H}$ and $\mathbf{A}_\mathrm{BS}^H\mathbf{G}\mathbf{A}_\mathrm{I}= \mathbf{\Lambda}_\mathbf{G}$. According to the sparse representation  of mm-wave channels in Section \ref{sparse},  $\mathbf{\Lambda}_\mathbf{G}$ and $\mathbf{\Lambda}_\mathbf{H}$ are sparse matrices with $P$ and $Q$ non-zero elements, respectively, which completes the proof of Lemma 2. $\hfill\blacksquare$ 

 Recall that Lemma 1 reveals the low-rank property of the channels themselves. In contrast, Lemma 2 indicates that there is another essential uniqueness, i.e., sparsity, inherented in the angular domain representations.
By exploiting Lemmas 1 and 2, we refine estimation problem \eqref{eqn:ls-func} as follows
\begin{equation}\label{eqn:formulate-l0}
\begin{array}{cl}
\displaystyle{\minimize_{ \hat{\mathbf{G}}, \hat{\mathbf{H}}}} &  \sum_{t=1}^{T}{\|\mathbf{r}_t - \hat{\mathbf{G}}\mathbf{\Phi}_t\hat{\mathbf{H}}\mathbf{s}_t\|^2} \\
\mathrm{subject \; to} &  \mathrm{rank}(\hat{\mathbf{G}}) = P, \quad \mathrm{rank}(\hat{\mathbf{H}}) = Q,\\
& \|\boldsymbol{\lambda}_{\hat{\mathbf{G}}}\|_0=P, \quad \|\boldsymbol{\lambda}_{\hat{\mathbf{H}}}\|_0=Q,
\end{array}
\end{equation}
where $\boldsymbol{\lambda}_{\hat{\mathbf{H}}}\triangleq\mathrm{vec}\left(\mathbf{A}_\mathrm{I}^H\hat{\mathbf{H}}\mathbf{A}_\mathrm{UE}\right)$ and $\boldsymbol{\lambda}_{\hat{\mathbf{G}}}\triangleq\mathrm{vec}\left(\mathbf{A}_\mathrm{BS}^H\hat{\mathbf{G}}\mathbf{A}_\mathrm{I}\right)$.
Unfortunately, problem \eqref{eqn:formulate-l0} is intractable due to the highly non-convex constraints. Hence, a globally optimal solution cannot be obtained in general.  To tackle the non-convex $\ell_0$-norm constraints  introduced by the low-rank and sparse properties of mm-wave channels, we resort to the $\ell_1$-norm regularization approach.  In particular, the cardinality constraint induced by the $\ell_0$-norm is relaxed by its convex envelop, i.e., the $\ell_1$-norm \cite{Berger2010}. In addition, the regularized terms are scaled by tuning parameters to avoid  over-fitting. Correspondingly, the estimation problem \eqref{eqn:formulate-l0} is reformulated as follows

% Nevertheless, a widely-used approach is to relax the $l_0$-norm constraints and  instead add its tightest convex approximation, i.e., the $\ell_1$-norm as penalty terms \cite{XLi2018, Berger2010}. Follow this idea, we resort to tackle the following problem
% Nevertheless, the theory of compressed sensing  reveals that the recovery of such sparse variables in \eqref{eqn:formulate-l0} can be achieved by minimizing their $\ell_1$-norm, i.e., the tightest convex approximation of $l_0$-norm \cite{Berger2010}. Therefore, we resort to tackle with the following problem instead of \eqref{eqn:formulate-l0}, whose  objective is in the same form to that of the well-known basis pursuit denoising (BPDN)
\begin{equation}\label{eqn:formulate-l1}
\begin{array}{cl}
\displaystyle{\minimize_{ \hat{\mathbf{G}}, \hat{\mathbf{H}}}}  &  f = \sum_{t=1}^{T}{\|\mathbf{r}_t - \hat{\mathbf{G}}\mathbf{\Phi}_t\hat{\mathbf{H}}\mathbf{s}_t\|^2} + \mu_\mathbf{G}\|\boldsymbol{\lambda}_{\hat{\mathbf{G}}}\|_1 + \mu_\mathbf{H}\|\boldsymbol{\lambda}_{\hat{\mathbf{H}}}\|_1 \\
\mathrm{subject \; to} &  \mathrm{rank}(\hat{\mathbf{G}}) = P, \quad \mathrm{rank}(\hat{\mathbf{H}}) = Q,
\end{array}
\end{equation}
% where the intractable $l_0$-norm constraints in \eqref{eqn:formulate-l0} are relaxed and the differentiable $l_1$-norms of $\boldsymbol{\lambda}_{\hat{\mathbf{G}}}$ and $\boldsymbol{\lambda}_{\hat{\mathbf{H}}}$ are correspondingly added to the objective as the penalty terms instead.
where $\mu_\mathbf{G}$ and $\mu_\mathbf{H}$ denote the tuning parameters that control the sparsity levels of $\boldsymbol{\lambda}_\mathbf{G}$ and $\boldsymbol{\lambda}_\mathbf{H}$, {\color{black}respectively}. 

% \emph{Remark 2:} In practical   $M$, $N_\mathrm{BS}$, and $N_\mathrm{UE}$ can not be infinitely large, and the assumption of Lemma 2 is not satisfied. Nevertheless, although $\boldsymbol{\lambda}_{\hat{\mathbf{H}}}$ and $\boldsymbol{\lambda}_{\hat{\mathbf{G}}}$ are not strictly sparse with small $M$, $N_\mathrm{BS}$, and $N_\mathrm{UE}$, it should be mentioned that their $\ell_1$-norms still concentrate on a few elements while other elements approach zero \cite{Fan2018}. In \cite{Tao2006}, the authors revealed that the recovery of such vectors can be achieved by minimizing their $\ell_1$-norm, which suggests the minimization of $\|\boldsymbol{\lambda}_{\hat{\mathbf{G}}}\|_1$ and $\|\boldsymbol{\lambda}_{\hat{\mathbf{H}}}\|_1$ in \eqref{eqn:formulate-l1} can still lead to good solution. 

Problem \eqref{eqn:formulate-l1} is still difficult to solve due to the coupled optimization variables in the objective function  $f$ and the highly non-convex low-rank constraints.  Therefore,  we   decouple the  optimization of the two variables in  problem (\ref{eqn:formulate-l1}) by applying the AM principle, which has been widely adopted in different wireless communication application scenarios, e.g.,  hybrid precoding  for massive MIMO systems \cite{yu_alternating_2016} and passive beamforming for IRS-assisted systems \cite{Guo:2020, Li:2020}. Specifically,  we first fix $\hat{\mathbf{H}}$ and minimize $f$ with respect to $\hat{\mathbf{G}}$. The  subproblem is given by
\begin{equation}\label{eqn:subpro-G}
\begin{array}{cl}
\displaystyle{\minimize_{\mathbf{X}}} &  f_1=\|\mathbf{R} - \mathbf{X}\mathbf{F}\|_F^2 + \mu_\mathbf{G}\|\boldsymbol{\lambda}_\mathbf{X}\|_1 \\
\mathrm{subject \; to} &   \mathrm{rank}(\mathbf{X}) = P, 
\end{array}
\end{equation}
where $\mathbf{R} = [\mathbf{r}_1, \ldots, \mathbf{r}_T]\in\mathbb{C}^{N_\mathrm{BS}\times T}$, $\mathbf{F}=[\mathbf{\Phi}_1\hat{\mathbf{H}}\mathbf{s}_1,\ldots, \mathbf{\Phi}_T\hat{\mathbf{H}}\mathbf{s}_T]\in\mathbb{C}^{M\times T}$, and $\mathbf{X}\triangleq\hat{\mathbf{G}}$ are defined for notational convenience. To address the non-convex fixed-rank constraint, we apply the MO technique to solve problem  (\ref{eqn:subpro-G})  in the following. 
\subsection{Preliminaries of MO}
By extending the definition of the real-valued fixed-rank manifold \cite{boumal2020intromanifolds} to the complex domain,  the feasible set of problem \eqref{eqn:subpro-G} can be represented as a typical Riemannian manifold \cite{He2020Cascaded}
\begin{equation}
\mathcal{M}_{P}\triangleq\left\{\mathbf{X} \in \mathbb{C}^{N_\mathrm{BS}\times M}: \operatorname{rank}(\mathbf{X})=P\right\},
\end{equation}
to which  optimization tools developed for the Euclidean space, e.g., the gradient descend and trust-region methods, can be transplanted \cite{boumal2020intromanifolds, absil2009optimization}. Before deriving the proposed algorithm, we first introduce  some key operations that are necessary for the Riemannian optimization method for $\mathcal{M}_P$.

\textit{1) Inner product:} By endowing the complex space $\mathbb{C}^{N_\mathrm{BS}\times M}$ with the Euclidean
metric, the standard inner product between two points $\mathbf{X}_{1}, \mathbf{X}_{2}\in \mathcal{M}_{P}$ is defined as follows
\begin{equation}
    \left\langle \mathbf{X}_{1}, \mathbf{X}_{2}\right\rangle=\Re\left\{\mathrm{tr}(\mathbf{X}_{1}^{H}\mathbf{X}_{2})\right\}.
\end{equation}

\textit{2) Tangent space:} For a point $\mathbf{X}\in \mathcal{M}_{P}$ on the manifold, its tangent space $T_\mathbf{X} \mathcal{M}_{P}$, which is
composed of all the vectors that tangentially pass
through $\mathbf{X}$, is  given by \cite{boumal2020intromanifolds}
\begin{equation}\begin{aligned}
T_\mathbf{X} \mathcal{M}_{P}
\triangleq\{\mathbf{X}_\mathrm{U} \mathbf{M}\mathbf{X}_\mathrm{V}^{H}+ \mathbf{U}_\mathrm{p} \mathbf{X}_\mathrm{V}^{H}+\mathbf{X}_\mathrm{U} \mathbf{V}_\mathrm{p}^{H}\},
\end{aligned}\end{equation}
where $\mathbf{X}_\mathrm{U}\in\mathbb{C}^{N_\mathrm{BS}\times P}$ and $\mathbf{X}_\mathrm{V}\in\mathbb{C}^{M\times P}$ denote the semi-unitary matrices containing the first $P$  left  and right singular vectors of $\mathbf{X}$, respectively. $\mathbf{M} \in \mathbb{C}^{P \times P}$ is an arbitrary matrix. In addition, $\mathbf{U}_\mathrm{p} \in \mathbb{C}^{N_\mathrm{BS} \times P}$ and $\mathbf{V}_\mathrm{p} \in \mathbb{C}^{M \times P}$ lie in the null spaces of $\mathbf{X}_\mathrm{U}$ and $\mathbf{X}_\mathrm{V}$, respectively, i.e., $\mathbf{U}_\mathrm{p}^{H} \mathbf{X}_\mathrm{U}= \mathbf{0}$ and $\mathbf{V}_\mathrm{p}^{H} \mathbf{X}_\mathrm{V}=\mathbf{0}$.

\textit{3) Orthogonal projection:}  The orthogonal projection of  $\mathbf{J}\in \mathbb{C}^{N_\mathrm{BS}\times M}$ onto the tangent space of $\mathbf{X}$, i.e., $T_\mathbf{X} \mathcal{M}_{P}$,  {\color{black}is} given by
\begin{equation}
    \label{tangent-pro}
    \mathrm{Proj}_{\mathbf{X}} (\mathbf{J})=  \mathbf{P}_{\mathbf{U}}\mathbf{J} \mathbf{P}_{\mathbf{V}}+\mathbf{P}_{\mathbf{U}}^{\perp} \mathbf{J}  \mathbf{P}_{\mathbf{V}}+\mathbf{P}_{\mathbf{U}} \mathbf{J}   \mathbf{P}_{\mathbf{V}}^{\perp},
\end{equation}
where $\mathbf{P}_\mathbf{U} =\mathbf{X}_\mathrm{U}\mathbf{X}_\mathrm{U}^H$, $\mathbf{P}_\mathbf{V} =\mathbf{X}_\mathrm{V}\mathbf{X}_\mathrm{V}^H$, $\mathbf{P}_{\mathbf{U}}^{\perp}=\mathbf{I}_{N_\mathrm{BS}} - \mathbf{P}_\mathbf{U}$, and $\mathbf{P}_{\mathbf{V}}^{\perp}=\mathbf{I}_{M} - \mathbf{P}_\mathbf{V}$ \cite{boumal2020intromanifolds}. 
\begin{figure}[!t]
 		\centering
 		\includegraphics[height=1.9in]{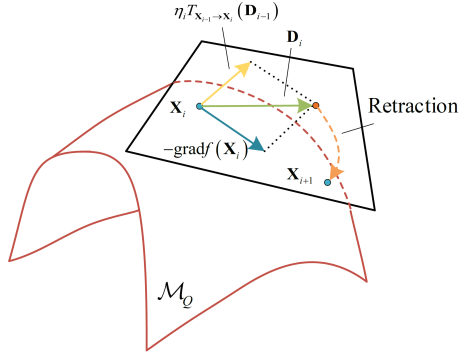}
 	\caption{Illustration of the generalized CG method for the fixed-rank manifold.}
 	    \label{manifold}
\end{figure}

\subsection{Conjugate Gradient Method on $\mathcal{M}_P$}
With the basic definitions of the key operations on $\mathcal{M}_P$ at hand, 
we can formulate the counterpart of the classic conjugate gradient (CG) algorithm in the Euclidean
space  on the manifold $\mathcal{M}_P$ \cite{boumal2020intromanifolds, absil2009optimization}. The main idea is illustrated in Fig. \ref{manifold}.  Specifically, in the $i$-th iteration initialized at the point $\mathbf{X}_i\in \mathcal{M}_P$, by introducing the definition of the linear space $T_{\mathbf{X}_i}\mathcal{M}_P$, the conventional CG algorithm applicable in the Euclidean space can  be applied to find a local minimizer  in the tangent space,  which is  subsequently mapped back to $\mathcal{M}_P$ to obtain $\mathbf{X}_{i+1}$. For problem \eqref{eqn:subpro-G},  the update rule of the search direction of the CG method in the tangent  space $T_{\mathbf{X}_i}$ is given by 
\begin{equation}
\label{direction}
    \mathbf{D}_i = -\mathrm{grad} f_1\left(\mathbf{X}_i\right) + \eta_iT_{\mathbf{X}_{i-1} \rightarrow \mathbf{X}_{i}}\left(\mathbf{D}_{i-1}\right),
\end{equation}
where the first term of \eqref{direction} is the negative Riemannian gradient representing the   steepest  descent  direction  of  the  objective  function $f_1$  in  the  tangent  space $T_{\mathbf{X}_i}\mathcal{M}_P$, and $\eta_i$ represents the chosen Polak-Ribiere parameter \cite[p. 42]{shewchuk1994introduction}. 
Since the conjugate direction in the previous iteration $\mathbf{D}_{i-1}$ does not lie in $T_{\mathbf{X}_i}\mathcal{M}_P$, the summation in \eqref{direction} can not be performed directly. To this end, we introduce the \textit{vector transport} operation to project $\mathbf{D}_{i-1}$ to the current tangent space $T_{\mathbf{X}_i}\mathcal{M}_P$. According to (\ref{tangent-pro}), the vector transport for $\mathcal{M}_P$ is given by
\begin{equation}
\begin{split}
 T_{\mathbf{X}_{i-1} \rightarrow \mathbf{X}_{i}}
 = \mathrm{Proj}_{\mathbf{X}_i} \left(\mathbf{D}_{i-1}\right).
 \end{split}
\end{equation}
Therefore, the remaining task to determine the conjugate direction in \eqref{direction} is to derive the Riemannian gradient.
%can be performed to find a local minimizer in $T_{\hat{\mathbf{H}}^{(i)}_{\mathrm{p}}} \mathcal{M}_{Q}$
%where $\alpha^{(i)}$ denotes the search step size  determined via Armijo backtracking \cite{MO3}.
%
%In order to find a local minimizer of problem (\ref{eqn:opt_prob1}), the Riemannian gradient representing the  steepest increase direction of the objective function in the tangent space needs to be determined. 
Since $\mathcal{M}_P$ is embedded in $\mathbb{C}^{N_\mathrm{BS}\times M}$, the Riemannian gradient is obtained by  projecting the  conjugate Euclidean gradient $\nabla_{\mathbf{X}_i^*} f_1$  onto the tangent space \cite{absil2009optimization}, i.e., 
\begin{equation}
\label{Riemannian-gradient}
    \mathrm{grad} f_1\left(\mathbf{X}_i\right)=  \mathrm{Proj}_{\mathbf{X}_i} \left(\nabla_{\mathbf{X}_i^*} f_1\right),
\end{equation}
where $\nabla_{\mathbf{X}_i^*} f_1$ is given by the following lemma.

\begin{lemma} \label{lemma2}
The  Euclidean  gradient of function $f_1$ with respect to $\mathbf{X}_i^*$ is given by
\begin{equation}\label{eqn:G-gradient}
    \nabla_{\mathbf{X}_i^*} f_1 = -\mathbf{R}\mathbf{F}^H + \mathbf{X}\mathbf{F}\mathbf{F}^H +  \frac{\mu_\mathbf{G}}{2}\mathbf{A}_\mathrm{BS}\mathbf{Y}\mathbf{A}_\mathrm{I}^H,
\end{equation}
where $\mathbf{Y}$ is computed as 
\begin{equation}\label{eqn:Y}
    [\mathbf{Y}]_{ij} = \frac{[\mathbf{A}_\mathrm{BS}^H\mathbf{X}\mathbf{A}_\mathrm{I}]_{ij}}{|[\mathbf{A}_\mathrm{BS}^H\mathbf{X}\mathbf{A}_\mathrm{I}]_{ij}|}.
\end{equation}
\end{lemma}
\textit{Proof}: Please refer to Appendix B. $\hfill\blacksquare$

Finally, based on the derived search direction $\mathbf{D}_i$,  an operation called \textit{retraction}  is introduced to find the destination on the manifold. Specifically, after moving forward along  the search direction $\mathbf{D}_i$  in the tangent space, we map the resulting point back to $\mathcal{M}_P$ itself  by solving the following optimization problem \cite{boumal2020intromanifolds}
\begin{equation}\label{eqn:retraction_problem}
    \mathcal{R}_{\mathbf{X}_i}(\kappa_i\mathbf{D}_i)={\arg \underset{\hat{\mathbf{X}} \in \mathcal{M}_P}{\min} }\quad\|\mathbf{X}_i+\kappa_i\mathbf{D}_i-\hat{\mathbf{X}}\|^{2},
\end{equation}
where $\kappa_i$ denotes the Armijo backtracking step size in the $i$-th iteration \cite[Eq. (59)]{shewchuk1994introduction}. A closed-form solution of problem \eqref{eqn:retraction_problem} can be  obtained via a truncated singular value decomposition (SVD)
\begin{equation}
\label{retract}
\mathcal{R}_{\mathbf{X}_i}\left(\kappa_i\mathbf{D}_i\right)= \sum_{i=1}^{P}\zeta_{i} \mathbf{u}_{i} \mathbf{q}_{i}^{H},
\end{equation}
where $\zeta_{i}$, $\mathbf{u}_{i}$, and ${\color{black}\mathbf{q}}_{i}$ are the ordered singular values, left singular vectors, and right singular vectors of $\mathbf{X}_i + \kappa_i\mathbf{D}_i$, respectively.
 The proposed generalized CG method for the fixed-rank manifold,  referred to as the \textbf{CG-MO algorithm},  is summarized in \textbf{Algorithm 1}, where $\epsilon_1$ is the convergence threshold. We note that the retraction in \eqref{retract} can be obtained via QR factorization \cite{boumal2020intromanifolds}. {\color{black}Hence}, the computational complexity of the \textbf{CG-MO algorithm} is given by $\mathcal{O}\left(\left(M +N_\mathrm{BS}\right)P^2\right)$.

\begin{algorithm}[t]
\label{alg:manifold}
	\caption{CG-MO Algorithm}
	\begin{algorithmic}[1]
	\STATE Randomly initialize $\mathbf{X}_0\in\mathcal{M}_P$ , set $i=0$ and $f_0=f_1\left(\mathbf{X}_0\right)$.
\REPEAT	
	\STATE  Compute the conjugate Euclidean gradient $\nabla_{\mathbf{X}_i^*} f_1$ according to \eqref{eqn:G-gradient};
	\STATE  Determine the Riemannian gradient $\mathrm{grad} f_1\left(\mathbf{X}_i\right)$ according to (\ref{Riemannian-gradient});
	\STATE  Choose Polak-Ribiere parameter $\eta_i$ \cite[p. 42]{shewchuk1994introduction} and obtain the conjugate search direction according to \eqref{direction}; 
	\STATE  Find $\mathbf{X}_{i+1}$ via  retraction \eqref{retract};
    \STATE  $i\leftarrow i+1$;
    \STATE $f_i=f_1\left(\mathbf{X}_{i}\right)$;
\UNTIL $f_{i-1}-f_i\le \epsilon_1$.
% 		\STATE Update $\mathbf{X}_{i}$ as the estimate of  ${\mathbf{G}}$.
	\end{algorithmic}
\end{algorithm}

\subsection{Estimation of $\mathbf{H}$}
In this subsection, we consider the optimization of $\hat{\mathbf{H}}$ for given $\hat{\mathbf{G}}$. The corresponding subproblem is obtained as follows
\begin{equation}\label{eqn:opt_prob2}
\begin{array}{cl}
\displaystyle{\minimize_{ \hat{\mathbf{H}}}} &  f_2 = \sum_{t=1}^{T}{\|\mathbf{r}_t - \hat{\mathbf{G}}\mathbf{\Phi}_t\hat{\mathbf{H}}\mathbf{s}_t\|^2} +  \mu_\mathbf{H}\|\boldsymbol{\lambda}_{\hat{\mathbf{H}}}\|_1 \\
\mathrm{subject \; to} &  \mathrm{rank}(\hat{\mathbf{H}}) = Q.
\end{array}
\end{equation}
Notice that the feasible set of problem \eqref{eqn:opt_prob2} is also a fixed-rank Riemannian manifold, i.e., $\mathcal{M}_{Q}\triangleq\left\{\mathbf{X} \in \mathbb{C}^{M\times N_\mathrm{UE}}: \operatorname{rank}(\mathbf{X})=Q\right\}$ and thus the \textbf{CG-MO algorithm} is also applicable. 
The main modification compared to the optimization of $\hat{\mathbf{G}}$ is the replacement of the conjugate Euclidean gradient in \eqref{eqn:G-gradient}  by the conjugate Euclidean gradient of $f_2$ with respect to $\hat{\mathbf{H}}$, which is given by
\begin{equation}
\label{g2}
    \nabla_{\hat{\mathbf{H}}^*}f_2 =\frac{\mu_{\hat{\mathbf{H}}}}{2}\mathbf{A}_\mathrm{I}\mathbf{Y}_2\mathbf{A}_\mathrm{UE}^H + \sum_{t=1}^T{\left(-\mathbf{\Phi}_t^H\hat{\mathbf{G}}^H\mathbf{r}_t\mathbf{s}^H_t+\mathbf{\Phi}_t^H\hat{\mathbf{G}}^H\hat{\mathbf{G}}\mathbf{\Phi}_t\hat{\mathbf{H}}\mathbf{s}_t\mathbf{s}_t^H\right)},
\end{equation} 
where $\mathbf{Y}_2$ is given by
\begin{equation}
    [\mathbf{Y}_2]_{ij} = \frac{[\mathbf{A}_\mathrm{I}^H\hat{\mathbf{H}}\mathbf{A}_\mathrm{UE}]_{ij}}{|[\mathbf{A}_\mathrm{I}^H\hat{\mathbf{H}}\mathbf{A}_\mathrm{UE}]_{ij}|}.
\end{equation}
 The derivation of \eqref{g2} is similar to that of \eqref{eqn:G-gradient}, and thus, {\color{black}it} is omitted here. The resulting overall estimation scheme  is referred to as the \textbf{MO-EST algorithm}, which is summarized in \textbf{Algorithm 2}, where $\epsilon_2$ is the convergence threshold. With the proposed algorithm, the objective values $f$ achieved by the sequence $\left\{\hat{\mathbf{H}}^{(k)}, \hat{\mathbf{G}}^{(k)}\right\}_{k\in \mathbb{N}}$ form a non-increasing sequence that converges to a stationary value, and any limit point of the sequence $\left\{\hat{\mathbf{H}}^{(k)}, \hat{\mathbf{G}}^{(k)}\right\}_{k\in \mathbb{N}}$ is a stationary point of problem \eqref{eqn:formulate-l1} \cite{absil2009optimization}. 

Although the proposed MO-EST algorithm has a lower computational complexity than the schemes  in \cite{wang_compressed_2020} and \cite{ma_joint_2020}, its computational complexity is still relatively high. Specifically, in each iteration of the MO  algorithm, the computational complexities of the conjugate gradient and the truncated SVD are $\mathcal{O}(MN_\mathrm{BS}(M + N_\mathrm{UE}+T))$ and $\mathcal{O}(MN^2_\mathrm{BS} +  MN^2_\mathrm{UE})$, respectively. Nevertheless,  as will be shown in Section VI, the MO-EST algorithm can serve as a  performance upper bound for channel estimation in IRS-assisted mm-wave MIMO systems. Besides, although it is difficult to analytically characterize the training overhead required for the MO-EST algorithm,  {\color{black}our simulation results in Section VI} show that the MO-EST algorithm significantly reduces the training overhead when compared with  conventional LS based channel algorithms \cite{lin_channel_2020, araujo_parafac-based_2020}.
\begin{algorithm}[t]
\caption{MO-EST Algorithm}
\label{MOAlg2}
\begin{algorithmic}[1]
% \REQUIRE $\mathbf{V}$, $\widetilde{\mathbf{Y}}_1$, $\widetilde{\mathbf{Y}}_2$
\STATE Randomly initialize $\hat{\mathbf{G}}^{(0)} \in \mathcal{M}_P$ and $\hat{\mathbf{H}}^{(0)} \in \mathcal{M}_Q$, set $k=0$, and $f^{(0)}=f\left(\hat{\mathbf{H}}^{(0)}, \hat{\mathbf{G}}^{(0)}\right)$.
\REPEAT	
\STATE $k\leftarrow k+1$;
\STATE Optimize $\hat{\mathbf{G}}^{(k)}$ for given $\hat{\mathbf{H}}^{(k-1)}$ by solving problem  (\ref{eqn:subpro-G}) with the CG-MO algorithm;
\STATE Optimize $\hat{\mathbf{H}}^{(k)}$ for given $\hat{\mathbf{G}}^{(k)}$ by solving problem  (\ref{eqn:opt_prob2}) with the CG-MO algorithm;
 \STATE $f^{(k)}=f\left(\hat{\mathbf{H}}^{(k)}, \hat{\mathbf{G}}^{(k)}\right)$;
\UNTIL $f^{(k-1)}-f^{(k)}\le \epsilon_2$; 
\STATE Update $\hat{\mathbf{G}}^{(k)}$ and $\hat{\mathbf{H}}^{(k)}$ as the estimates of ${\mathbf{G}}$ and ${\mathbf{H}}$.
\end{algorithmic}\label{alg:manifold}
\end{algorithm}

%=======================================
%               Sec4: CS-EST
%=======================================

\section{Proposed CS-EST Algorithm}
\label{sec:OMP}
To further reduce the computational complexity of  channel estimation,  in this section, we propose a low-complexity CS-based  algorithm as an alternative. Recall that  the sparse mm-wave channel representation in Section II-B has three key components, namely the AoAs, AoDs, and the complex gains. Therefore, different from the MO-EST algorithm that directly estimates the channel matrices, the proposed CS-based algorithm recovers the channel matrices by estimating the {\color{black}three} associated  key components instead.
\subsection{Conventional CS-Based Approach}
\begin{figure}[!t]
 		\centering
 		\includegraphics[height=1.0in]{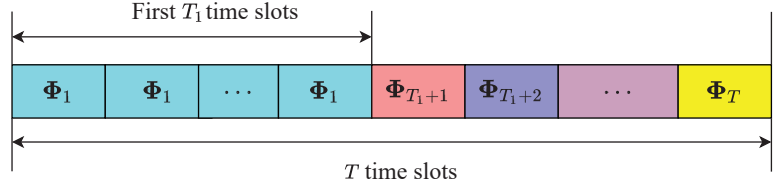}
 	\caption{ Reflection matrices adopted during the training phase for the CS-EST algorithm.}
 	    \label{fig:CS_protocol}
 	\end{figure}

The received pilot {\color{black}signal} in the $t$-th time slot  in \eqref{eqn:rt} can be rewritten as
\begin{equation} \label{eqn:rewrite-rt}
      \mathbf{r}_{t} = (\mathbf{s}_t^T\otimes \mathbf{I}_{N_\mathrm{BS}})(\mathbf{H}^T\odot\mathbf{G})\mathbf{v}_t + \mathbf{z}_t,
\end{equation}
where we exploited $\mathrm{vec}(\mathbf{A}\mathbf{B}\mathbf{C}) = (\mathbf{C}^T\otimes\mathbf{A})\mathrm{vec}(\mathbf{B})$ and $\mathrm{vec}(\mathbf{A}\mathbf{D}\mathbf{C}) = (\mathbf{C}^T\odot\mathbf{A})\mathbf{d}$, which hold for arbitrary matrices $\mathbf{A}$, $\mathbf{B}$, $\mathbf{C}$. Furthermore,  $\mathbf{d}$ denotes the vector of diagonal elements of an arbitrary diagonal matrix $\mathbf{D}$, i.e., $\mathbf{D} = \mathrm{diag}(\mathbf{d})$. Substituting the angular domain representation  \eqref{eqn:VAD} into \eqref{eqn:rewrite-rt}, we have
\begin{equation} \label{eqn:ybt-cs2}
\begin{split}
      \mathbf{r}_{t}    &  = (\mathbf{s}_t^T\otimes \mathbf{I}_{N_\mathrm{BS}})\left(\left(\mathbf{A}_\mathrm{UE}^*\mathbf{\Lambda}_\mathbf{H}^T\mathbf{A}_\mathrm{I}^T\right)\odot\left({\mathbf{A}}_\mathrm{BS}{\mathbf{\Lambda}}_\mathbf{G}{\mathbf{A}}_\mathrm{I}^H\right)\right)\mathbf{v}_t + \mathbf{z}_{t} \\
      & \overset{(a)}{=}   (\mathbf{s}_t^T\otimes \mathbf{I}_{N_\mathrm{BS}})(\mathbf{A}_\mathrm{UE}^*\otimes{\mathbf{A}}_\mathrm{BS})(\mathbf{\Lambda}_\mathbf{H}^T\otimes{\mathbf{\Lambda}}_\mathbf{G})(\mathbf{A}_\mathrm{I}^T\odot{\mathbf{A}}_\mathrm{I}^H)\mathbf{v}_t + \mathbf{z}_{t}\\
      & =  \underbrace{(\mathbf{v}_t^T\otimes\mathbf{s}_t^T\otimes \mathbf{I}_{N_\mathrm{BS}})\left((\mathbf{A}_\mathrm{I}^T\odot{\mathbf{A}}_\mathrm{I}^H)^T\otimes(\mathbf{A}_\mathrm{UE}^*\otimes{\mathbf{A}}_\mathrm{BS})\right)}_{\boldsymbol{\Psi}_t}\underbrace{\mathrm{vec}(\mathbf{\Lambda}_\mathbf{H}^T\otimes{\mathbf{\Lambda}}_\mathbf{G})}_{\boldsymbol{\mu}} + \mathbf{z}_{t},
      \end{split}
\end{equation}
where $(a)$ follows from $(\mathbf{A}\mathbf{B})\odot(\mathbf{C}\mathbf{D})=(\mathbf{A}\otimes\mathbf{C})(\mathbf{B}\odot\mathbf{D})$ and $\boldsymbol{\mu}\in\mathbb{C}^{G_\mathrm{UE}G_\mathrm{BS}G_\mathrm{I}^2}$ is a sparse vector with $PQ$ non-zero elements. According to \eqref{eqn:ybt-cs2}, the overall received pilots in $T$ successive time slots, i.e., $\hat{\mathbf{r}}=[\mathbf{r}_1^T,\ldots, \mathbf{r}_T^T]^T$ can be rewritten as
$\hat{\mathbf{r}} = \mathbf{\Psi}\boldsymbol{\mu} + \hat{\mathbf{z}},$
where $\mathbf{\Psi} = [\boldsymbol{\Psi }_1^T,\ldots, \boldsymbol{\Psi }_T^T]^T$ and $\hat{\mathbf{z}}=[\mathbf{z}_1^T,\ldots, \mathbf{z}_T^T]^T$. Thus,  the recovery of $\boldsymbol{\mu}$ is a classical sparse signal recovery problem, which can be formulated as follows
\begin{equation}\label{eqn:5}
\begin{array}{cl}
\displaystyle{\minimize_{\boldsymbol{\mu}}} &  \| \hat{\mathbf{r}}-\mathbf{\Psi}\boldsymbol{\mu}\|^2 \\
\mathrm{subject \; to} &  \|\boldsymbol{\mu}\|_0 = PQ.
\end{array}
\end{equation}
 CS-based algorithms, e.g., the GAMP and OMP methods, can be directly applied to recover $\boldsymbol{\mu}$, which, however,   leads to a prohibitively high computational complexity of $\mathcal{O}({TG_\mathrm{I}^2G_\mathrm{UE}G_\mathrm{BS}N_\mathrm{BS}})$ \cite{ma_joint_2020} and a training overhead of $\mathcal{O}({PQ\mathrm{log}(G_\mathrm{UE}G_\mathrm{BS}G_\mathrm{I}^2}))$ \cite{wang_compressed_2020}.  Therefore, in order to reduce the computational complexity, we propose to separate the overall estimation phase into three stages, where in each stage  a low-complexity CS method can be applied. Specifically, in the first  stage, we estimate the AoAs at the BS  based on the observations in the first $T_1$ time slot, while  the AoDs at the UE are estimated in the second stage by collecting the received pilots in all $T$ time slots. Finally, in the third stage, we estimate the channel gains and recover the cascaded channel based on the results of the previous two stages.  The proposed estimation protocol is illustrated in Fig. \ref{fig:CS_protocol}. Specifically, we set   the reflection matrix  $\mathbf{\Phi}_t$ {\color{black}such that it is} fixed as $\mathbf{\Phi}_1$ during the first $T_1$ time slots but changes in each of the other time slots, which facilitates {\color{black}the formulation of} the estimation in each stage  as a  sparse signal recovery problem.  
\subsection{Estimation of the  AoDs at the UE}
In the first stage, we estimate the AoDs at the UE. Based on the angular domain representation in Section \ref{sparse} and \eqref{eqn:rt},  the received pilots in the first $T_1$ time slots can be rewritten as {\color{black}follows}
\begin{equation} \label{eqn:ybt-cs}
\begin{split}
      \mathbf{r}_t^H = \mathbf{s}_t^H\mathbf{H}\mathbf{\Phi}_1^H\mathbf{G}^H + \mathbf{z}_t^H
      =\mathbf{s}_t^H\mathbf{A}_\mathrm{UE}\underbrace{\mathbf{\Lambda}_\mathbf{H}^H\mathbf{A}_\mathrm{I}^H\mathbf{\Phi}_1^H\mathbf{G}^H}_{\mathbf{\Gamma}_{\mathrm{UE}}} + \mathbf{z}_t^H, \quad   t\in\{1,2,\ldots, T_1\}. 
\end{split}
\end{equation}
Since $\mathbf{\Lambda}_{\mathbf{H}}$ is a sparse matrix with $Q$ non-zero elements,  $\mathbf{\Gamma}_{\mathrm{UE}}\in\mathbb{C}^{G_\mathrm{UE}\times N_\mathrm{BS}}$ is  a row-sparse matrix with $Q$ non-zero rows.  By collecting the pilots received in the first $T_1$ time slots, we have
\begin{equation}\label{eqn:y2_eqn}
    \mathbf{R}_1^H=\underbrace{\mathbf{S}_1\mathbf{A}_\mathrm{UE}}_{{\mathbf{\Theta }}}\mathbf{\Gamma}_\mathrm{UE} + \mathbf{Z}_1^H,
\end{equation}
where $\mathbf{R}_1= [\mathbf{r}_1,\ldots, \mathbf{r}_{T_1}]\in\mathbb{C}^{N_\mathrm{BS}\times T_1}$, $\mathbf{S}_1 = [\mathbf{s}_1,\ldots, \mathbf{s}_{T_1}]^H\in\mathbb{C}^{T_1\times N_\mathrm{UE}}$, and $\mathbf{Z}_1=[\mathbf{z}_1,\ldots, \mathbf{z}_{T_1}]\in\mathbb{C}^{N_\mathrm{BS}\times T_1}$.   Notice that the $i$-th  row of $\mathbf{\Gamma}_{\mathrm{UE}}$ is non-zero only when the $i$-th column of $\mathbf{A}_\mathrm{UE}$ corresponds to one of the  transmit antenna array response vectors $\mathbf{a}_{\mathrm{UE}}\left(\psi_{\mathrm{t}}^{q}\right)$,  $q=1, \ldots, Q$, of the reflecting channel $\mathbf{H}$ defined in \eqref{channel_model}. Thus, the estimation task in the first stage is to identify the  $Q$ columns of ${\mathbf{A}}_\mathrm{UE}$ corresponding to non-zero rows of   $\mathbf{{\Gamma}}_{\mathrm{UE}}$  based on the observation matrix $\mathbf{R}_1$ in \eqref{eqn:y2_eqn}, which is equivalent to the estimation of the AoDs at the UE.  This is a sparse recovery problem {\color{black}that  can be solved} with the OMP algorithm \cite{Heath2014}.  The relevant pseudo code of the OMP algorithm is provided in \textbf{Algorithm 3}, where $\mathbf{\Theta }^{(k)}$ and $\mathbf{A}_\mathrm{UE}^{(k)}$ respectively denote the $k$-th column of $\mathbf{\Theta }$ and the $k$-th column of $\mathbf{A}_\mathrm{UE}$. In each iteration, we find {\color{black}the} column of $\mathbf{A}_\mathrm{UE}$  most closely related to the updated residual. After $Q$ iterations, an $N_\mathrm{UE}\times Q$ matrix $\bar{\mathbf{A}}_\mathrm{UE}$ can be constructed by extracting the $Q$ columns of ${\mathbf{A}}_\mathrm{UE}$ corresponding to the non-zero rows of $\mathbf{\Lambda}_\mathbf{H}^H$.

\begin{algorithm}[t]
\label{alg:manifold}
	\caption{OMP Algorithm for the First Stage}
	\begin{algorithmic}[1]
	\STATE Initialize $\bar{\mathbf{\Theta}}$ and $\bar{\mathbf{A}}_\mathrm{UE}$ as  empty matrices, $\mathbf{R}_{\mathrm{res}} = \mathbf{R}_1$, and $\mathbf{\Theta }=\mathbf{S}_1\mathbf{A}_\mathrm{UE}$.
	\FOR{$q\le Q$}
	\STATE $\boldsymbol{\Psi}=\mathbf{\Theta }^{H}\mathbf{R}_{\text {res }}$;
	\STATE $k=\arg \max _{\ell}\left(\boldsymbol{\Psi} \boldsymbol{\Psi}^{H}\right)_{\ell, \ell}$;
	\STATE $\bar{\mathbf{\Theta }}=\left[\bar{\mathbf{\Theta }} \mid \mathbf{\Theta }^{(k)}\right]$;
	\STATE $\bar{\mathbf{A}}_\mathrm{UE}=\left[\bar{\mathbf{A}}_\mathrm{UE} \mid \mathbf{A}_\mathrm{UE}^{(k)}\right]$;
	\STATE $\mathbf{\Gamma}_{\mathrm{UE}}=\left(\bar{\mathbf{\Theta }}^{H} \bar{\mathbf{\Theta }}\right)^{-1} \bar{\mathbf{\Theta }}^{H} \mathbf{R}_{\text {res }}$;
	\STATE $\mathbf{R}_{\mathrm{res}}=\mathbf{R}_1 - \bar{\mathbf{\Theta }}\mathbf{\Gamma}_{\mathrm{UE}}$;
	\ENDFOR
	\STATE Output $\bar{\mathbf{A}}_\mathrm{UE}$ as the reduction of ${\mathbf{A}}_\mathrm{UE}$ for the third stage.
	\end{algorithmic}
\end{algorithm}

\subsection{Estimation of the  AoAs at the BS}
Similarly, in the second stage, we estimate the AoAs at the BS. The received pilots in the $t$-th time slot can be rewritten as {\color{black}follows}
\begin{equation}
    \mathbf{r}_t = \mathbf{A}_\mathrm{BS}\mathbf{\Lambda}_\mathbf{G}\mathbf{A}_\mathrm{I}^H\mathbf{\Phi}_t\mathbf{H}\mathbf{s}_t + \mathbf{z}_t,
\end{equation}
and then the  pilots received in all $T$ time slots can be modeled as
\begin{equation} \label{eqn:ybt-cs}
      \mathbf{R} =\mathbf{A}_\mathrm{BS}\underbrace{\mathbf{\Lambda}_\mathbf{G}\mathbf{A}_\mathrm{I}^H\mathbf{F}}_{\boldsymbol{\Gamma}_{\mathrm{BS}}}  + \mathbf{Z},\\
\end{equation}
where $\mathbf{R}$, $\mathbf{F}$, and $\mathbf{Z}$ are defined as in  \eqref{eqn:subpro-G}. As $\mathbf{\Gamma}_{\mathrm{BS}}\in\mathbb{C}^{G_\mathrm{BS}\times T}$ is also a row-sparse matrix, by applying the OMP algorithm, we can obtain an $N_\mathrm{BS}\times P$ matrix $\bar{{\mathbf{A}}}_\mathrm{BS}$ whose columns correspond to   the non-zero rows of $\mathbf{\Lambda}_\mathbf{G}$. Analogously, the $P$ columns of $\bar{{\mathbf{A}}}_\mathrm{BS}$ are associated with the receive
antenna array response vectors $\mathbf{a}_{\mathrm{r}}\left(\theta_{\mathrm{r}}^{p}\right),$  $p=1, \ldots, P,$ defined in \eqref{channel_model}, which is equivalent to the  estimation of the AoAs at the BS.

\subsection{Estimation of the Cascaded Channel}
Based on the estimated $\bar{\mathbf{A}}_\mathrm{BS}$ and $\bar{\mathbf{A}}_\mathrm{UE}$ in the  previous two stages, the angular domain representation of $\mathbf{G}$ and $\mathbf{H}$ can be reduced to $\mathbf{G}=\bar{\mathbf{A}}_\mathrm{BS}\bar{\mathbf{\Lambda}}_\mathbf{G}\mathbf{A}_\mathrm{I}^H$ and $\mathbf{H}={\mathbf{A}}_\mathrm{I}\bar{\mathbf{\Lambda}}_\mathbf{H}\bar{\mathbf{A}}_\mathrm{UE}^H$, where $\bar{\mathbf{\Lambda}}_\mathbf{G}\in\mathbb{C}^{P\times G_\mathrm{I}}$ and $\bar{\mathbf{\Lambda}}_\mathbf{H}\in\mathbb{C}^{G_\mathrm{I}\times Q}$ are two submatrices  obtained by eliminating  all zero rows of $\mathbf{\Lambda}_\mathbf{G}$ and  all zero columns of $\mathbf{\Lambda}_\mathbf{H}$, respectively.  The expression for $\mathbf{r}_t$ can then be rewritten as {\color{black}follows} 
\begin{equation}\label{y3t:eqn}
\begin{split}
    \mathbf{r}_{t}  & = \bar{\mathbf{A}}_\mathrm{BS}\bar{\mathbf{\Lambda}}_\mathbf{G}\mathbf{A}_\mathrm{I}^H\boldsymbol{\Phi}_{t}{\mathbf{A}}_\mathrm{I}\bar{\mathbf{\Lambda}}_\mathbf{H}\bar{\mathbf{A}}_\mathrm{UE}^H\mathbf{s}_{t}+ \mathbf{z}_t\\
    &=\left(\left(\mathbf{s}_{t}^T\bar{\mathbf{A}}_\mathrm{UE}^*\right)\otimes\bar{\mathbf{A}}_\mathrm{BS}\right)\left(\bar{\mathbf{\Lambda}}_\mathbf{H}^T\otimes\bar{\mathbf{\Lambda}}_\mathbf{G}\right)({\mathbf{A}}_\mathrm{I}^T\odot\mathbf{A}_\mathrm{I}^H)\mathbf{v}_{t} + \mathbf{z}_t,
    \end{split}
\end{equation}
where we used $\left(\mathbf{A}\mathbf{B}\right)\odot\left(\mathbf{C}\mathbf{D}\right)=\left(\mathbf{A}\otimes\mathbf{C}\right)\left(\mathbf{B}\odot\mathbf{D}\right).$ Based on the formulation of the overcomplete matrices in \eqref{eqn:UPA_dictionary}, we have that $\mathbf{A}_\mathrm{I}^T\circ \left(\mathbf{a}_j^H\otimes \mathbf{1}_{G_\mathrm{I}}\right)=\mathbf{L}_j\mathbf{A}_\mathrm{I}^T$, where $\mathbf{a}_j\in\mathbb{C}^M$ denotes the $j$-th column of $\mathbf{A}_\mathrm{I}$ and $\mathbf{L}_j\in\mathbb{C}^{G_\mathrm{I}\times G_\mathrm{I}}$ is a permutation matrix that rearranges the rows of $\mathbf{A}_\mathrm{I}$ \cite{2017Zhangxianda}. Therefore, the term  $\left(\bar{\mathbf{\Lambda}}_\mathbf{H}^T\otimes\bar{\mathbf{\Lambda}}_\mathbf{G}\right)({\mathbf{A}}_\mathrm{I}^T\odot\mathbf{A}_\mathrm{I}^H)$ in \eqref{y3t:eqn} can be simplified as follows
\begin{equation}\label{hat_lambda:eqn}
\begin{split}
   \left(\bar{\mathbf{\Lambda}}_\mathbf{H}^T\otimes\bar{\mathbf{\Lambda}}_\mathbf{G}\right)({\mathbf{A}}_\mathrm{I}^T\odot\mathbf{A}_\mathrm{I}^H) &= \left[ {\begin{array}{*{20}{c}}
{\bar{\boldsymbol{\lambda}}_1\otimes\bar{\mathbf{\Lambda}}_\mathbf{G}}&{\ldots}&{\bar{\boldsymbol{\lambda}}_{G_\mathrm{I}}\otimes\bar{\mathbf{\Lambda}}_\mathbf{G}}
\end{array}} \right] \left[ {\begin{array}{*{20}{c}}
{\mathbf{L}_1\mathbf{A}_\mathrm{I}^T}\\
{\vdots}\\
{\mathbf{L}_{G_\mathrm{I}}\mathbf{A}_\mathrm{I}^T}\\
\end{array}} \right]\\
&=\underbrace{\left(\sum_{j=1}^{G_\mathrm{I}}{\left(\bar{\boldsymbol{\lambda}}_j\otimes\bar{\mathbf{\Lambda}}_\mathbf{G}\right)\mathbf{L}_j}\right)}_{\hat{\mathbf{\Lambda}}}\mathbf{A}_\mathrm{I}^T,
\end{split}
\end{equation}
where $\bar{\boldsymbol{\lambda}}_j\in\mathbb{C}^Q$ denotes the $j$-th column of $\bar{\mathbf{\Lambda}}_\mathbf{H}^T$ and $\hat{\mathbf{\Lambda}}\in\mathbb{C}^{PQ\times G_\mathrm{I}}$ is a sparse matrix with $PQ$ non-zero elements. Finally, by substituting the result of \eqref{hat_lambda:eqn} into \eqref{y3t:eqn}, we have
\begin{equation}
    \mathbf{r}_{t}=\left(\left(\mathbf{s}_{t}^T\bar{\mathbf{A}}_\mathrm{UE}^*\right)\otimes\bar{\mathbf{A}}_\mathrm{BS}\right)\hat{\mathbf{\Lambda}}{\mathbf{A}}_\mathrm{I}^T\mathbf{v}_{t} + \mathbf{z}_{t}
    =\left(\left(\mathbf{v}_{t}^T\mathbf{A}_\mathrm{I}\right)\otimes\left(\mathbf{s}_{t}^T\bar{\mathbf{A}}_\mathrm{UE}^*\right)\otimes\bar{\mathbf{A}}_\mathrm{BS}\right)\boldsymbol{\lambda}+ \mathbf{z}_{t},\nonumber
\end{equation}
where ${\boldsymbol{\lambda}}=\mathrm{vec}(\hat{\mathbf{\Lambda}})\in \mathbb{C}^{PQG_\mathrm{I}}$ is a sparse vector with $PQ$ non-zero elements.  Subsequently, all   pilots received over all $T$ time slots can be collected in vector
\begin{equation}\label{block3}
    \hat{\mathbf{r}} = \left[{\begin{array}{*{20}{c}}
\left(\mathbf{v}_{1}^T\mathbf{A}_\mathrm{I}\right)\otimes\left(\mathbf{s}_{1}^T\bar{\mathbf{A}}_\mathrm{UE}^*\right)\otimes\bar{\mathbf{A}}_\mathrm{BS}\\
\vdots\\
\left(\mathbf{v}_{T}^T\mathbf{A}_\mathrm{I}\right)\otimes\left(\mathbf{s}_{T}^T\bar{\mathbf{A}}_\mathrm{UE}^*\right)\otimes\bar{\mathbf{A}}_\mathrm{BS} 
\end{array}} \right]\boldsymbol{\lambda}+ \hat{\mathbf{z}}.
\end{equation}
  For the sake of  complexity  reduction, we adopt the OMP algorithm to recover the  sparse vector $\boldsymbol{\lambda}$.   Finally,  with the obtained $\boldsymbol{\lambda}$, the  the cascaded channel $\mathbf{H}_\mathrm{c}\triangleq \mathbf{H}^T\odot\mathbf{G}$ can be recovered as {\color{black}follows}
\begin{equation}\label{eqn:39}
    {\mathbf{H}}_\mathrm{c}=(\bar{\mathbf{A}}_\mathrm{UE}^*\bar{\mathbf{\Lambda}}_\mathbf{H}^T\mathbf{A}_\mathrm{I}^T)\odot(\bar{\mathbf{A}}_\mathrm{BS}\bar{\mathbf{\Lambda}}_\mathbf{G}\mathbf{A}_\mathrm{I}^H) = (\bar{\mathbf{A}}_\mathrm{UE}^*\otimes\bar{\mathbf{A}}_\mathrm{BS})\bar{\boldsymbol{\Lambda}}\mathbf{A}_\mathrm{I}^T,
\end{equation}
where $\bar{\boldsymbol{\Lambda}}$ is obtained by reshaping $\boldsymbol{\lambda}$ to a $PQ\times G_\mathrm{I}$ matrix and the proposed estimation scheme is referred to as the \textbf{CS-EST algorithm}. Although the proposed approach cannot separately estimate the individual channel matrices, i.e., $\mathbf{G}$ and $\mathbf{H}$, in the next section,  we will illustrate that  knowledge of the cascaded channel matrix $\mathbf{H}_\mathrm{c}$ is sufficient for   system design after the channel estimation phase, e.g., for downlink beamforming design. The computational {\color{black}complexities} of the three stages of the proposed CE-EST algorithm {\color{black}are} $\mathcal{O}(QT_1N_\mathrm{UE}G_\mathrm{UE})$, $\mathcal{O}(PTN_\mathrm{BS}G_\mathrm{BS})$, and $\mathcal{O}({P^2Q^2TG_\mathrm{I}N_\mathrm{BS}})$, respectively, which is significantly {\color{black}lower} compared with that {\color{black}for} directly solving problem \eqref{eqn:ybt-cs} and {\color{black}that of} the MO-EST algorithm  proposed in Section III. On the other hand, note that recovering an $n\times 1$ vector with $m$ non-zero elements requires the dimension of the observation to be on  the order of $\mathcal{O}(m\mathrm{log}(mn))$  \cite{wang_compressed_2020, 2017Zhangxianda}.  Therefore, the {\color{black}required} training overhead of the proposed CS-EST algorithm is given by $T\ge\mathcal{O}(Q\mathrm{log}(QG_\mathrm{UE})+PQ\mathrm{log}(PQG_\mathrm{I}))$, which is typically much smaller than that of  simple LS estimation, i.e., $T\ge MN_\mathrm{UE}$, without  consideration of the unique properties of mm-wave channels. The training overhead reduction shall also be verified via simulation in Section VI.

\section{ALT-WMMSE Algorithm}
In this section, we propose a downlink beamforming design based on the knowledge of {\color{black}the} cascaded channel matrix,  $\mathbf{H}_\mathrm{c}$, obtained with the channel estimation schemes proposed in Sections III and IV. 
\subsection{Problem Formulation}
By utilizing the reciprocity of the uplink and downlink channels, the estimated uplink channel matrix can be utilized for downlink data transmission in  the IRS-assisted system considered in Section II. Assuming that the BS aims to transmit $N_\mathrm{s}$ data streams to the UE,  the  received signal $\mathbf{y}\in\mathbb{C}^{N_\mathrm{UE}}$ can be written as follows
\begin{equation}\label{eqn:downlink}
    \mathbf{y} = \mathbf{H}^H\mathbf{\Phi}_\mathrm{d}\mathbf{G}^H\mathbf{F}\mathbf{x} + \mathbf{z}_\mathrm{d},
\end{equation}
where  $\mathbf{F}\in\mathbb{C}^{N_\mathrm{BS}\times N_\mathrm{s}}$ represents the beamformer  at the BS and $\mathbf{x}\in\mathbb{C}^{N_\mathrm{s}}$ denotes the  symbol vector with $\mathbb{E}\{\mathbf{x}\mathbf{x}^H\}=\mathbf{I}_{N_\mathrm{s}}$ without loss of generality. $\mathbf{\Phi}_\mathrm{d}=\mathrm{diag}(\mathbf{v}_\mathrm{d})$ denotes the downlink reflection matrix and $\mathbf{z}_\mathrm{d}\in\mathbb{C}^{N_\mathrm{UE}}$ denotes the downlink additive Gaussian noise with $\mathbf{z}_\mathrm{d}\sim\mathcal{CN}(\mathbf{0}, \sigma^2_\mathrm{d}\mathbf{I}_{N_\mathrm{UE}})$.   By exploiting a property of the Khatri-Rao product, we have
  \begin{equation}\label{eqn:He}
     \mathbf{H}^H\mathbf{\Phi}_\mathrm{d}\mathbf{G}^H=\mathrm{mat}\left((\mathbf{G}^*\odot\mathbf{H}^H)\mathbf{v}_\mathrm{d}\right)=\mathrm{mat}(\mathbf{K}\mathbf{H}_\mathrm{c}^*\mathbf{v}_\mathrm{d})\triangleq\mathbf{H}_\mathrm{e},
  \end{equation}
 where $\mathrm{mat}(\cdot)$ denotes the operation that reshapes an $N_\mathrm{BS}N_\mathrm{UE}\times 1$ vector  to an $N_\mathrm{UE}\times N_\mathrm{BS}$ matrix. $\mathbf{K}\in\mathbb{C}^{N_\mathrm{BS}N_\mathrm{UE}\times N_\mathrm{BS}N_\mathrm{UE}}$ denotes the commutation matrix, which is a  constant matrix for given $N_\mathrm{BS}$ and $N_\mathrm{UE}$ \cite{2017Zhangxianda}. $\mathbf{H}_\mathrm{e}$ is  defined as the effective channel matrix, based on which \eqref{eqn:downlink} can be further rewritten as $\mathbf{y} = \mathbf{H}_\mathrm{e}\mathbf{F}\mathbf{x} + \mathbf{z}_\mathrm{d}.$   Then, the achievable spectral efficiency when the transmitted symbols follow a complex Gaussian distribution can be expressed as {\color{black}follows}
\begin{equation}\label{eqn:Rate}
    R=\left(1-\frac{T}{T_\mathrm{tot}}\right)\log |\mathbf{I}_{N_\mathrm{s}}+\frac{1}{\sigma^2_\mathrm{d}}\mathbf{F}^H\mathbf{H}_\mathrm{e}^H\mathbf{H}_\mathrm{e}\mathbf{F}|,
\end{equation}
where $T_\mathrm{tot}$ denotes the total number of time slots within the channel coherence time. In this section, our goal is to maximize the spectral efficiency  by optimizing the beamformer at the BS and the reflection matrix at the IRS.

\emph{Remark 2:} From (42) and (43), {\color{black}we observe that} the CSI of the effective channel, $\mathbf{H}_\mathrm{e}$, is sufficient for the design of the downlink beamformer and {\color{black}the} reflection matrix. {\color{black}With} the MO-EST algorithm, we obtain the CSI for $\mathbf{H}$ and $\mathbf{G}$ separately. Hence, $\mathbf{H}_\mathrm{e}$ can be directly composed according to \eqref{eqn:He}. On the other hand, {\color{black}with} the CS-EST algorithm, the cascaded channel, $\mathbf{H}_\mathrm{c}$, is estimated via \eqref{eqn:39}, and therefore, $\mathbf{H}_\mathrm{e}$ can be formed according to \eqref{eqn:He}. In the following, we show how to design the beamformer and {\color{black}the} IRS reflection matrix for  downlink data transmission  based on $\mathbf{H}_\mathrm{e}$, or, in other words, the CSI estimated by the two proposed algorithms.

To this end, we resort to  the equivalent WMMSE minimization problem \cite[eq. (32)]{2008WMMSE}
\begin{equation}\label{prb:wmmse-pro}
\begin{array}{cl}
\displaystyle{\minimize_{{{\bf{F}}},  {{\bf{W}}}, \mathbf{v}_\mathrm{d}, {\boldsymbol{\Omega}}}} & {g= {{\text{tr}}\left( {{{\boldsymbol{\Omega}}} {{{\bf{ E}}}}} \right) - \log \left| {\boldsymbol{\Omega}} \right|} }  \\
\mathrm{subject \; to} &\left\| \mathbf{F} \right\|_F^2 \le 1, \quad 
|[\mathbf{v}_\mathrm{d}]_n|=1 \quad \forall n,
\end{array}
\end{equation}
where $\left\| \mathbf{F}\right\|_F^2 \le 1$ denotes the normalized transmit power constraint and $|[\mathbf{v}_\mathrm{d}]_n|=1$ is the constant modulus constraint imposed by the  phase shifters. Furthermore, $\mathbf{E}\in\mathbb{C}^{N_\mathrm{s}\times N_\mathrm{s}}$ is the MSE matrix, which is given by \cite{lin_hybrid_2019}
\begin{equation}\label{eqn:E}
{\mathbf{E}} = \mathbb{E}\left[ {\left( {{\mathbf{x}} -{\mathbf{y}}} \right){{\left( {{\mathbf{x}} - {\mathbf{y}}} \right)}^H}} \right]
= {{\mathbf{I}}_{{N_{\text{s}}}}} - {\mathbf{F}}^H{\mathbf{H}_\mathrm{e}}^H{{\mathbf{W}}} - {\mathbf{W}}^H{{\mathbf{H}_\mathrm{e}}}{{\mathbf{F}}} + \sigma^2_\mathrm{d}{\mathbf{W}}^H{{\mathbf{W}}} + {\mathbf{W}}^H{{\mathbf{H}_\mathrm{e}}}{{\mathbf{F}}}{\mathbf{F}}^H{\mathbf{H}_\mathrm{e}^H}{{\mathbf{W}}},\nonumber
\end{equation}
with $\mathbf{\Omega}\in\mathbb{C}^{N_\mathrm{s}\times N_\mathrm{s}}$ and $\mathbf{W}\in\mathbb{C}^{N_\mathrm{UE}\times N_\mathrm{s}}$ being two auxiliary  variables.  By applying the AM principle, in the following, we {\color{black}alternately} optimize these variables. 

\subsection{Optimization of the Beamformer}
Aiming at  problem \eqref{prb:wmmse-pro}, when fixing $\mathbf{F}$ and $\mathbf{v}_\mathrm{d}$,  closed-form solutions for $\mathbf{W}$ and $\mathbf{\Omega}$ are given by
\begin{equation}\label{eqn:WD-opt}
    \mathbf{W}=(\mathbf{H}_\mathrm{e}\mathbf{F}\mathbf{F}^H\mathbf{H}_\mathrm{e}^H + \sigma^{2}_\mathrm{d}\mathbf{I}_{N_\mathrm{UE}} )^{-1}\mathbf{H}_\mathrm{e}\mathbf{F},\quad
    \mathbf{\Omega} = \mathbf{E}^{-1}.
\end{equation}
Similarly, when $\mathbf{v}_\mathrm{d}$, $\mathbf{W}$, and $\mathbf{\Omega}$ are fixed, a closed-form solution of $\mathbf{F}$  is given by
\begin{equation}\label{eqn:F-BB}
    \mathbf{F} = \xi\Tilde{\mathbf{F}},\quad \xi = \left(\operatorname{tr}\left( \Tilde{\mathbf{F}} \Tilde{\mathbf{F}}^{H} \right)\right)^{-\frac{1}{2}},
\end{equation}
where 
     $\Tilde{\mathbf{F}}=\left( \mathbf{H}_\mathrm{e}^H\mathbf{W} \boldsymbol{\Omega} \mathbf{W}^{H}\mathbf{H}_\mathrm{e}+\sigma^{2} \psi \mathbf{I}_{N_\mathrm{BS}}\right)^{-1}  \mathbf{H}_\mathrm{e}^H\mathbf{W} \boldsymbol{\Omega}$ denotes the unnormalized baseband beamformer
and $\psi \triangleq \operatorname{tr}\left(\mathbf{\Omega} \mathbf{W}^{H} \mathbf{W}\right)$. Hence, {\color{black}recalling}    the definition of $\mathbf{H}_\mathrm{e}$ {\color{black}in \eqref{eqn:He}}, the CSI of $\mathbf{H}_\mathrm{c}$ is sufficient to obtain  closed-form solutions. 
\subsection{Optimization of the Reflection Matrix}
In this subsection, we focus on the optimization of $g$ {\color{black}in \eqref{prb:wmmse-pro}} with respect to $\mathbf{v}_\mathrm{d}$ by fixing the other variables. The corresponding subproblem  is given by
\begin{equation}\label{prb:wmmse-pro-v}
\begin{array}{cl}
\displaystyle{\minimize_{ \mathbf{v}_\mathrm{d}}} & g_1(\mathbf{v}_\mathrm{d})=\mathrm{tr}(- \mathbf{\Omega}{\mathbf{F}}^H{\mathbf{H}_\mathrm{e}^H}{{\mathbf{W}}} - \mathbf{\Omega}\mathbf{W}^H\mathbf{H}_\mathrm{e}\mathbf{F} + \mathbf{\Omega}{\mathbf{W}}^H{{\mathbf{H}_\mathrm{e}}}{{\mathbf{F}}}{\mathbf{F}}^H{\mathbf{H}^H_\mathrm{e}}{{\mathbf{W}}})  \\
\mathrm{subject \; to} &  
|[\mathbf{v}_\mathrm{d}]_n|=1 \quad \forall n.
\end{array}
\end{equation}
Instead of directly solving problem \eqref{prb:wmmse-pro-v}, we further substitute the closed-form solution of $\mathbf{W}$ given by \eqref{eqn:WD-opt}  into $g_1(\mathbf{v}_\mathrm{d})$, {\color{black}which yields}
\begin{equation}\label{g1:pro}
    g_1(\mathbf{v}_\mathrm{d}) = \mathrm{tr}\left(\left(\mathbf{\Omega}^{-1} + \frac{1}{\sigma^2_\mathrm{d}}\mathbf{\Omega}^{-1}\mathbf{F}^H\mathbf{H}_\mathrm{e}^H\mathbf{H}_\mathrm{e}\mathbf{F}\right)^{-1}\right).
\end{equation}
The feasible set of $\mathbf{v}_\mathrm{d}$ is also a well-known complex circle Riemannian manifold \cite{yu_alternating_2016, lin_hybrid_2019}, i.e., $\mathcal{M}\triangleq\{\mathbf{x} \in \mathbb{C}^{M}: |[\mathbf{x}]_n|=1, \forall n \}.$
Therefore, the MO technique can  be applied to optimize $\mathbf{v}_\mathrm{d}$. Based on  basic differentiation rules for complex-valued matrices \cite{2020ComplexDerive}, the differential is given by
\begin{equation}\label{eqn:differentiate}
        \mathrm{d}(g_1)  \mathop  = \limits^{(b)}  -\frac{1}{\sigma_\mathrm{d}^2}\mathrm{tr}\left(\mathbf{T}^{-2}\mathbf{\Omega}^{-1}\mathbf{F}^H\mathrm{d}\left(\mathbf{H}_\mathrm{e}^H\right)\mathbf{H}_\mathrm{e}\mathbf{F}\right)
         \mathop  = \limits^{(c)}  -\frac{1}{\sigma^2_\mathrm{d}}\mathbf{m}^T\mathrm{d}\left(\mathbf{H}_\mathrm{e}^H\right)
         \mathop  = \limits^{(d)}  -\frac{1}{\sigma^2_\mathrm{d}}\mathbf{m}^T\mathbf{H}_\mathrm{c}\mathrm{d}(\mathbf{v}_\mathrm{d}^*),
\end{equation}
where $\mathbf{T} \triangleq\mathbf{\Omega}^{-1} + \frac{1}{\sigma^2_\mathrm{d}}\mathbf{\Omega}^{-1}\mathbf{F}^H\mathbf{H}_\mathrm{e}^H\mathbf{H}_\mathrm{e}\mathbf{F}$ and $\mathbf{m}\triangleq\mathrm{vec}\left((\mathbf{H}_\mathrm{e}\mathbf{F}\mathbf{T}^{-2}\mathbf{\Omega}^{-1}\mathbf{F}^H)^T\right)$ are defined for notational brevity. Note that $(b)$ follows from $\mathrm{d}(\mathbf{X}^{-1})=-\mathbf{X}^{-1}\mathrm{d}(\mathbf{X}^{-1})\mathbf{X}^{-1}$,  $(c)$ follows from  $\mathrm{tr}(\mathbf{A}\mathbf{B}) = \left(\mathrm{vec}\left(\mathbf{A}^T\right)\right)^T\mathrm{vec}(\mathbf{B})$, and $(d)$ follows from $  \mathrm{vec}(\mathbf{H}_\mathrm{e}^H)=(\mathbf{H}^T\odot\mathbf{G})\mathbf{v}_\mathrm{d}^*$. Based on \eqref{eqn:differentiate} and using $\mathrm{d}(g_1)=\left(\nabla_{\mathbf{v}_\mathrm{d}^*} g_1\right)^T\mathrm{d}(\mathbf{v}_\mathrm{d}^*)$, the  gradient of $g_1$ with respect to $\mathbf{v}_\mathrm{d}^*$ is given by
\begin{equation}\label{eqn:Euc-gradient}
    \nabla_{\mathbf{v}_\mathrm{d}^*} g_1 = -\frac{1}{\sigma^2_\mathrm{d}}\mathbf{H}_\mathrm{c}^T\mathbf{m}.
\end{equation}
 Given the derived conjugate Euclidean  gradient,  MO can be straightforwardly applied to optimize $\mathbf{v}_\mathrm{d}$ under the constant modulus constraint \cite{yu_alternating_2016}. {\color{black}From \eqref{eqn:differentiate} and \eqref{eqn:Euc-gradient} we observe that}  knowledge of $\mathbf{H}_\mathrm{c}$ is also sufficient for the optimization of $\mathbf{v}_\mathrm{d}$.

\subsection{ALT-WMMSE Algorithm}
Based on the previous subsections, {\color{black}the} joint optimization of the {\color{black}beamformer} and the reflection matrix {\color{black}can be accomplished} by alternately optimizing the variables.  {\color{black}The overall ALT-WMMSE algorithm is summarized in \textbf{Algorithm 4}},  where $\epsilon_3$ is the convergence threshold. Notice that since in each step the variables are updated based on  closed-form solutions or  monotonous descent algorithms \cite{boumal2020intromanifolds},  the ALT-WMMSE algorithm is guaranteed to converge to a stationary point of problem \eqref{prb:wmmse-pro} \cite{boumal2020intromanifolds}. The  computational complexity of the proposed algorithm is $\mathcal{O}(N_\mathrm{BS}N_\mathrm{UE}M)$, which is mainly caused by the computation of the Euclidean gradient.

\begin{algorithm}[t]
\label{alg:manifold}
	\caption{ALT-WMMSE Algorithm}
	\begin{algorithmic}[1]
	\STATE Set $i=0$,  randomly initialize $\mathbf{v}_\mathrm{d}^{(0)}$ and   $\mathbf{F}^{(0)}$.
	\STATE Solve $\mathbf{W}^{(0)}$ and $\mathbf{\Omega}^{(0)}$ according to \eqref{eqn:WD-opt} and set $g^{(0)}=g(\mathbf{W}^{(0)}, \mathbf{F}^{(0)}, \mathbf{\Omega}^{(0)}, \mathbf{v}_\mathrm{d}^{(0)})$;
\REPEAT	
	\STATE  Update $\mathbf{v}_\mathrm{d}^{(i+1)}$ via MO optimization with the Euclidean gradient derived in \eqref{eqn:Euc-gradient};
	\STATE  Update $\mathbf{W}^{(i+1)}$ and $\mathbf{\Omega}^{(i+1)}$ with the closed-form solutions in \eqref{eqn:WD-opt};
	\STATE  Update $\mathbf{F}^{(i+1)}$ with the closed-form solutions in \eqref{eqn:F-BB};
    \STATE  $i\leftarrow i+1$;
    \STATE $g^{(i)}=g(\mathbf{W}^{(i)}, \mathbf{F}^{(i)}, \mathbf{\Omega}^{(i)}, \mathbf{v}_\mathrm{d}^{(i)})$;
\UNTIL $g^{(i-1)} - g^{(i)}\le \epsilon_3$.
	\end{algorithmic}
\end{algorithm}

\section{Simulation Results}
\subsection{Simulation Setup}
 According to the channel model in (\ref{channel_model}),  without loss of generality, we denote $p = 1$ and $q = 1$ as the indices of the LoS components in $\mathbf{G}$ and $\mathbf{H}$, respectively. If not specified otherwise, for both $\mathbf{H}$ and $\mathbf{G}$, the same number of paths {\color{black}is} assumed, i.e., $P=Q\triangleq K=3$. {\color{black}The complex channel gains of the LoS components are distributed as $\alpha_1\sim \mathcal{CN}(0,\tau_{\mathrm{BI}})$ and $\beta_1\sim \mathcal{CN}(0,\tau_\mathrm{IU})$, while those of the NLoS components are distributed as  $\alpha_i\sim \mathcal{CN}(0,10^{-0.5}\tau_{\mathrm{BI}})$ and $\beta_i\sim \mathcal{CN}(0,10^{-0.5}\tau_\mathrm{IU})$ for $i= 2,\ldots, K$ \cite{zhou2021channel, Peilan_channel}, where $\tau_\mathrm{BI}$ and $\tau_\mathrm{IU}$ are  given by \cite{Wang2021Rate}
 \begin{equation}
     \tau_\mathrm{BI} = 10^{-6.14 - 2\log_\mathrm{10}(d_\mathrm{BI})}\quad \mathrm{and} \quad \tau_\mathrm{IU} = 10^{-6.14 - 2\log_\mathrm{10}(d_\mathrm{IU})}.
 \end{equation}
 Here, $d_\mathrm{BI}$ represents the distance between the BS and the IRS and is set to $150$ m, while  $d_\mathrm{IU}$ denotes the distance between the IRS and the UE and is set to $10$ m}.  The azimuth and elevation AoAs/AoDs are generated uniformly distributed in $(0$, $2\pi]$.  $T_\mathrm{tot}$ is set to $2000$, corresponding to a channel coherence time of $5$ ms {\color{black}for} a transmission bandwidth of {\color{black}$4\times 10^5$} Hz.   The convergence thresholds $\epsilon_1$, $\epsilon_2$, and $\epsilon_3$   are all set to  $10^{-3}$. {\color{black}The uplink training pilot-to-noise-ratio (PNR) is defined as $\frac{P_\mathrm{tr}\tau_\mathrm{BI}\tau_\mathrm{IU}}{\sigma^2}$, while the downlink transmission signal-to-noise-ratio (SNR) is defined as $\frac{\tau_\mathrm{BI}\tau_\mathrm{IU}}{\sigma_\mathrm{d}^2}$}.
 All simulation results are averaged over $1000$ independent channel realizations.

% To show the effectiveness of the proposed algorithms, the LS-based PARAFAC algorithm  \cite{araujo_parafac-based_2020, Wei2020ParaFAC} and the conventional GAMP algorithm \cite{wang_compressed_2020} are adopted as two benchmarks. For our proposed TS-GAMP algorithm, we set the angular resolutions of all pre-defined dictionaries as $G_\mathrm{I}=G_\mathrm{UE}=G_\mathrm{BS}=256$. However, in order to achieve an affordable complexity, the resolutions of the dictionaries of the conventional GAMP algorithm, is set as $G_\mathrm{I}=G_\mathrm{UE}=G_\mathrm{BS}=64$. Even so, as analyzed in Section IV, the complexity of the proposed TS-GAMP algorithm is still far lower than the GAMP algorithm.

\subsection{Performance of the ALT-WMMSE Algorithm}

\begin{figure}
\centering
	\includegraphics[width=2.7in]{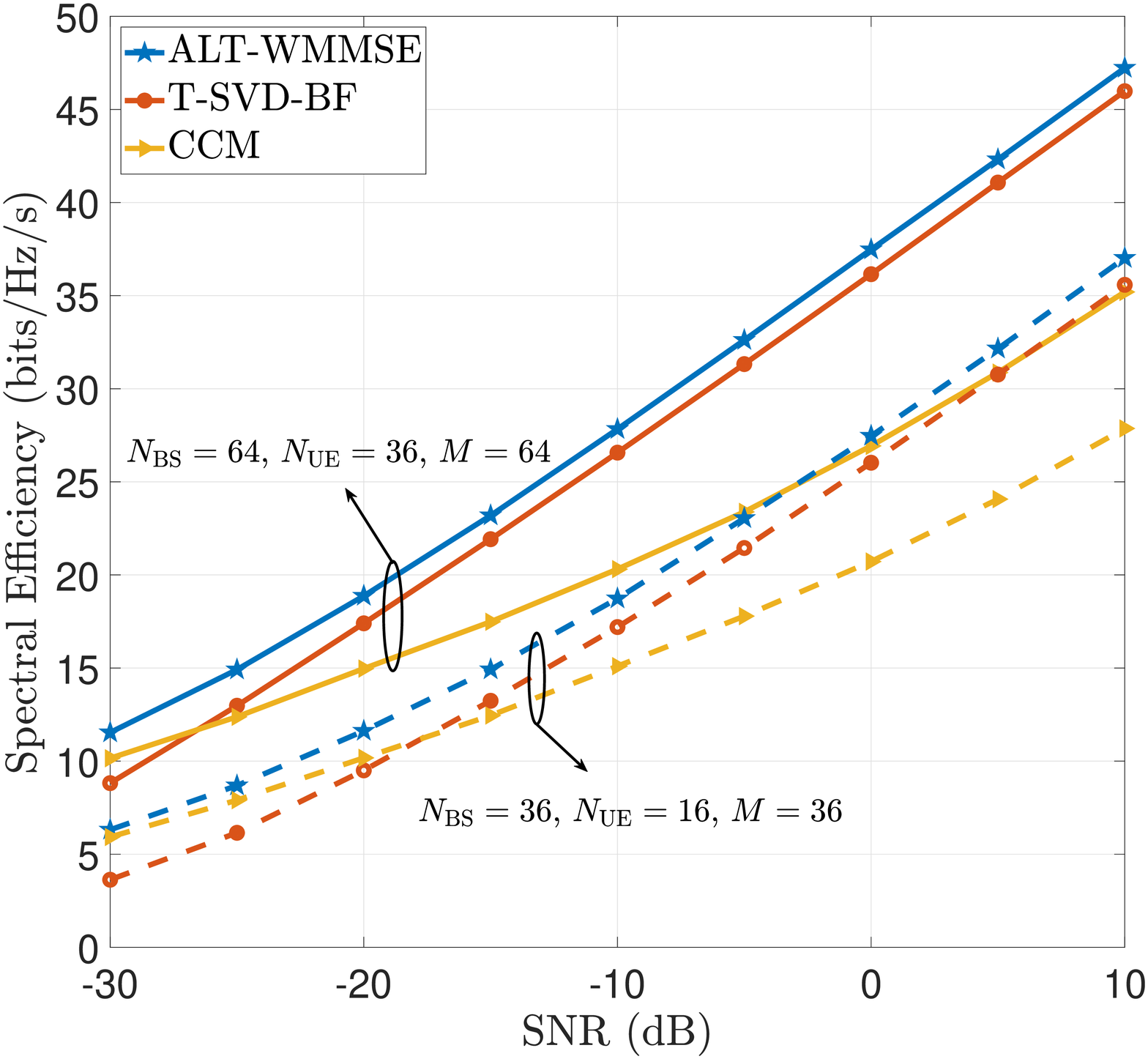}
 	\caption{Spectral efficiency versus SNR for different beamforming algorithms  in IRS-assisted mm-wave MIMO systems.}
 	    \label{Rate1}
 	    \vspace{-10pt}
\end{figure}

In this subsection, to show the effectiveness of the proposed ALT-WMMSE beamforming algorithm, we first evaluate {\color{black}its} performance  in terms of spectral efficiency  \eqref{eqn:Rate} assuming perfect knowledge of $\mathbf{H}$ and $\mathbf{G}$ with $T=0$. Two state-of-the-art passive beamforming algorithms are adopted as  benchmarks, namely, the complex circle manifold (CCM) algorithm \cite{Pan2020Rate} and the truncated-SVD-based beamforming (T-SVD-BF) algorithm \cite{Wang2021Rate}.  The CCM algorithm optimizes $\mathbf{v}_\mathrm{d}$ by minimizing the objective of problem \eqref{prb:wmmse-pro-v} without substituting the optimal solution \eqref{eqn:WD-opt} into \eqref{prb:wmmse-pro-v}. On the other hand, the T-SVD-BF algorithm assumes  equal power allocation among different data streams to simplify the problem. Fig. \ref{Rate1} shows the spectral efficiency as a function of the SNR for two different system configurations when $N_\mathrm{s}=3$. As can be observed, the proposed ALT-WMMSE algorithm achieves the highest spectral efficiency over   the entire SNR range considered. This shows the superiority of the proposed ALT-WMMSE beamforming algorithm for rate maximization. In particular, as the optimal solution of $\mathbf{W}$ is used for solving problem \eqref{prb:wmmse-pro-v}, the spectral efficiency achieved by the ALT-WMMSE algorithm is significantly improved compared to  the CCM algorithm. Furthermore, since  equal power allocation is only asymptotically optimal in  the high SNR regime, the performance   of the T-SVD-BF algorithm entails a significant loss for low SNRs. Therefore, in the remainder  of this section,  the ALT-WMMSE {\color{black}algorithm} is adopted for beamforming  to evaluate the downlink spectral efficiency with imperfect  knowledge of $\mathbf{H}_\mathrm{c}$, which is obtained with different channel estimation schemes.

\subsection{Benchmark Schemes}
In this section, the PARAFAC algorithm  \cite{araujo_parafac-based_2020, Wei2020ParaFAC} and the GAMP algorithm \cite{wang_compressed_2020} are adopted as two state-of-the-art benchmark {\color{black}schemes}. The PARAFAC algorithm models the received pilots as a tensor and then obtains the channel matrices based on the LS criterion without exploiting the low-rank and sparse properties of mm-wave channels  revealed in Lemmas 1 and 2.  The GAMP algorithm treats the channel estimation problem as a classical sparse signal recovery problem \eqref{eqn:5}. During the channel training  phase, for all considered algorithms, we adopt random quasi-omnidirectional training beams \cite{XLi2018}, i.e., the entries of $\mathbf{v}_t$ and $\mathbf{s}_t$,  $t=1,\ldots, T$, are randomly chosen on the complex unit circle, and set $N_\mathrm{BS}=36$, $N_\mathrm{UE}=16$, and $M=36$. To facilitate a fair comparison between  the CS-EST and GAMP algorithms,  we list the  number of required floating point operations (FLOPs)  with respect to {\color{black}the} angular resolution, $F$, for both algorithms in Table I when $T=100$. In particular, the angular resolutions at the BS and the UE are set to $G_\mathrm{BS}=G_\mathrm{UE}=F$. Since the IRS is typically considered {\color{black}to be} a large-scale planar array, a higher resolution is assumed at the IRS, i.e., $G_\mathrm{I}=4F$, to achieve a satisfactory estimation performance.\par
\setlength{\abovecaptionskip}{-1pt}
\begin{table}[]
\caption{ Number of  {\color{black}FLOPs required} for the CS-EST and GAMP algorithms with respect to $F$.}\label{Tab:FLOPs}
\centering
\begin{tabular}{|c|c|c|c|c|c|}
\hline
\diagbox{Algorithm}{Angular resolution $F$}       & 8        & 16        & 32        & 64        & 128       \\ \hline
CS-EST & $9.35\times 10^6$ & $1.87\times 10^7$  & $3.74\times 10^7$  & $7.48\times 10^7$  & $1.5\times 10^8$  \\ \hline
GAMP   & $4.72\times 10^9$ & $7.54\times 10^{10}$ & $1.21\times 10^{12}$ & $1.93\times 10^{13}$ & $3.09\times 10^{14}$ \\ \hline
\end{tabular}
\end{table}

Consistent with the complexity analysis in Sections III and IV, Table \ref{Tab:FLOPs} shows {\color{black}that} the  required numbers of FLOPs for the CS-EST  and GAMP algorithms scales with $F$ and $F^4$, respectively. Thus, in the remainder of this section, for  the proposed CS-EST algorithm, we adopt  $F=64$, i.e., $G_\mathrm{BS}=64$, $G_\mathrm{UE}=64$, and $G_\mathrm{I}=256$ ($16\times 16$). In contrast, in order to achieve an affordable complexity, for the GAMP algorithm $F$ is reduced to $F=16$, i.e., $G_\mathrm{BS}=16$, $G_\mathrm{UE}=16$, and $G_\mathrm{I}=64$ ($8\times 8$). Note that the computational complexity of the proposed CS-EST algorithm is still 
much lower than that of the GAMP algorithm even though a higher angular resolution is adopted for the CS-EST algorithm. 
\setlength{\abovecaptionskip}{-7pt}
\begin{figure}[t]
\centering
\begin{minipage}[t]{0.49\textwidth}
\centering
\centering
 		\includegraphics[width=2.7in]{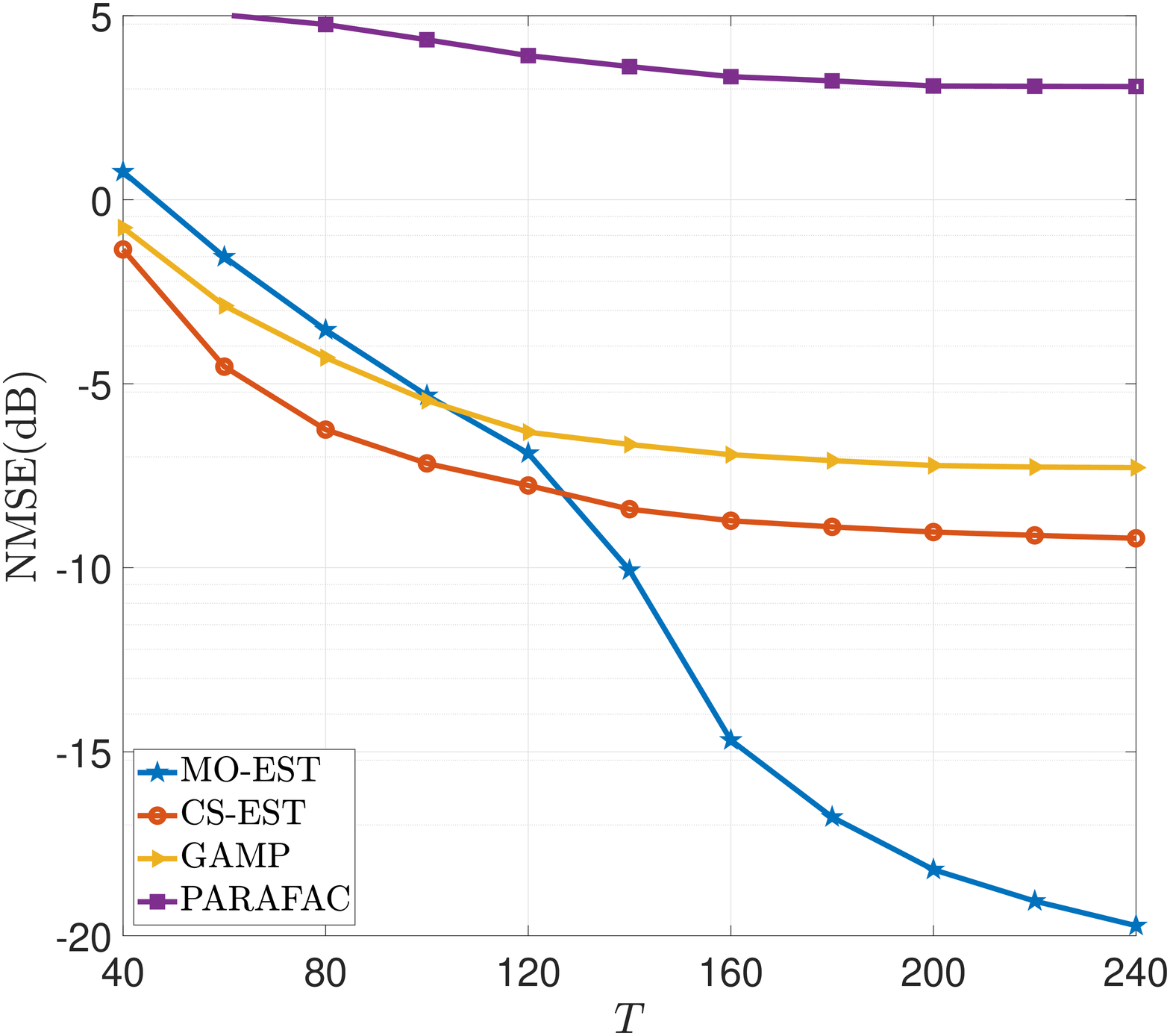}
 	\caption{NMSE versus $T$ for different channel estimation algorithms when $\mathrm{PNR}=0$ dB.}
 	    \label{NMSEvsT}
\end{minipage}
\begin{minipage}[t]{0.49\textwidth}
\centering
 		\includegraphics[width=2.7in]{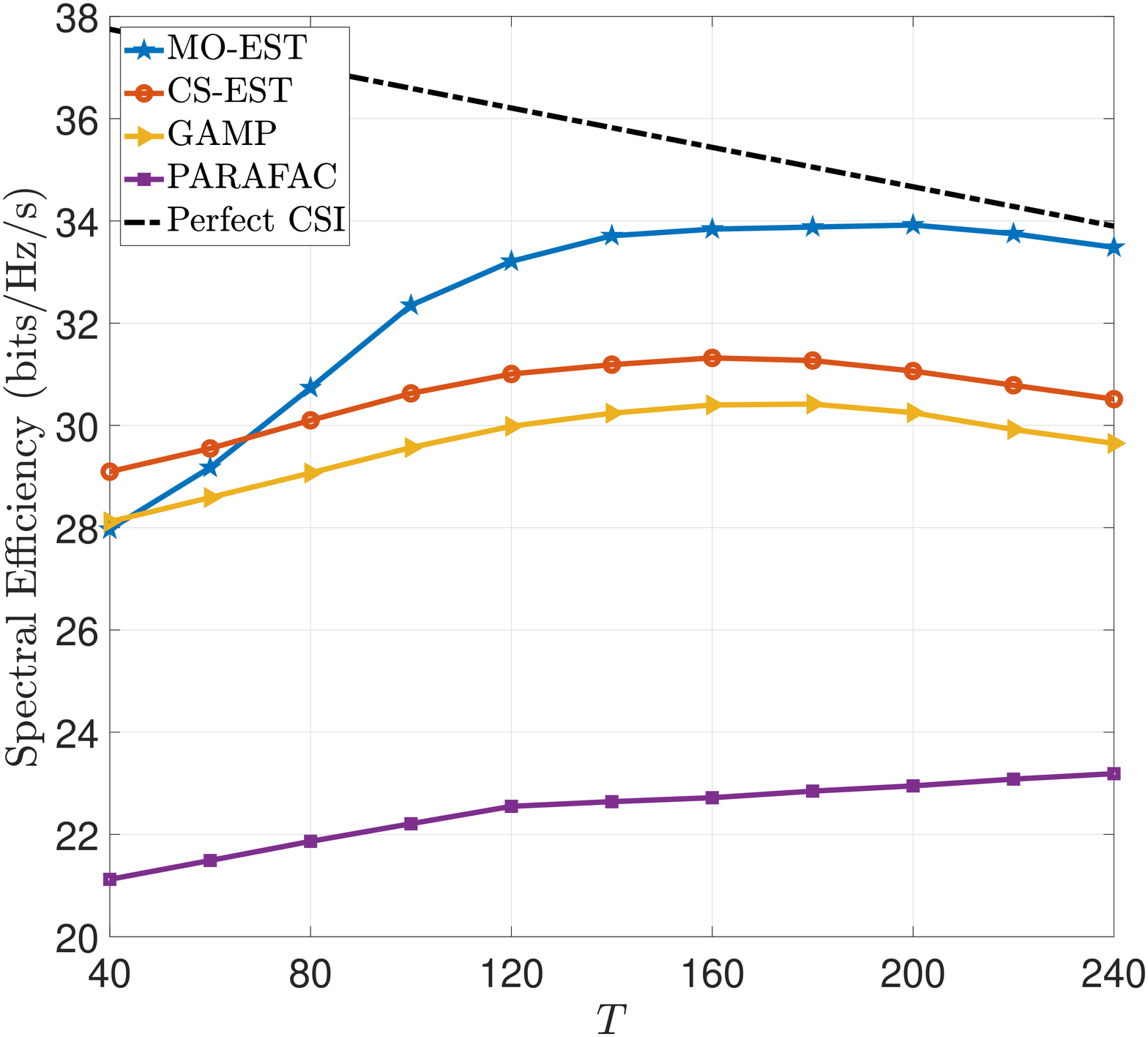}
 	\caption{Spectral efficiency versus $T$ for {\color{black}different} channel estimation algorithms when $\mathrm{PNR}=\mathrm{SNR}=10$ dB.}
 	    \label{TvsRate}
\end{minipage}
\end{figure}
\subsection{System Performance Versus  Training Overhead}
Now, we evaluate the performance of the two proposed channel estimation algorithms.
In Fig. \ref{NMSEvsT}, we  plot the normalized mean square error (NMSE) versus the training overhead,  $T$, {\color{black}for} $\mathrm{PNR}=0$ dB. The NMSE is defined as $\mathbb{E}\left\{\|\mathbf{H}_\mathrm{c} - \hat{\mathbf{H}}_\mathrm{c}\|_F^2 / \|\mathbf{H}_\mathrm{c}\|_F^2\right\}$. As can be observed, the NMSE of all algorithms decreases with increasing training overhead $T$. Since the PARAFAC algorithm requires a training overhead of $T\ge MN_\mathrm{UE}$ \cite{araujo_parafac-based_2020}, {\color{black}it is  not  well suited} for the considered range of $T$ and  results in the worst performance. In contrast, thanks to the fixed-rank constraints and $\ell_1$-norm regularization, MO is able  to find a sparse solution in the angular domain and therefore  achieves a lower NMSE with much fewer pilots. Besides, although the MO-EST algorithm suffers from {\color{black}a} performance loss for small $T$, its performance improves rapidly as $T$ increases and it achieves the lowest NMSE  among all investigated algorithms for $T\ge 130$. On the other hand, as shown in Section IV, the number of time slots required for the CS-EST algorithm is only in the order of $\mathcal{O}(Q\mathrm{log}(QG_\mathrm{UE})+PQ\mathrm{log}(PQG_\mathrm{I}))$.  Consequently,  the CS-EST algorithm achieves the lowest NMSE for small $T$. Meanwhile, it outperforms the GAMP algorithm as it can afford a  higher angular resolution, i.e., larger $G_\mathrm{BS}$, $G_\mathrm{UE}$, and $G_\mathrm{I}$, {\color{black}because of its} lower computational complexity according to Table I.  However,  the finite angular resolution becomes the main bottleneck for further improving of the performance for large $T$, which causes the NMSEs of both the CS-EST and the GAMP algorithms to saturate. 

% \footnote{As  $\mathbf{G}$ and $\mathbf{H}$ are coupled in the received signals, there inevitably {\color{black}exists  a scaling ambiguity} between $\hat{\mathbf{G}}$ and $\hat{\mathbf{H}}$.  Besides, in Section V, we illustrated that the downlink beamforming design only depends on the cascaded channel $\mathbf{H}_\mathrm{c}$. Therefore, the NMSE of  $\hat{\mathbf{H}}_\mathrm{c}$ is adopted as  performance metric  to avoid the scaling issues.}

 In Fig. \ref{TvsRate}, we plot the spectral efficiency versus the training overhead, $T$.
 For beamforming,  the ALT-WMMSE algorithm is applied, where  the effective channel estimate $\hat{\mathbf{H}}_\mathrm{e}$ obtained by different algorithms is adopted. We set  downlink $\mathrm{SNR}=10$ dB, $N_\mathrm{s}=3$, and $\mathrm{PNR}=10$ dB. As can be observed from Fig. \ref{TvsRate}, the spectral efficiency of all algorithms increases with  $T$ when $T\le 160$, thanks to the improved accuracy of the CSI. {\color{black}If $T$ is increased further} (e.g., $T\ge 200$), for the GAMP, CS-EST, and MO-EST algorithms, {\color{black}the larger $T$ cannot significantly improve the CSI further, and thus, the corresponding spectral efficiencies}  decrease due to the increasing training overhead. The MO-EST algorithm achieves the best performance when $T\ge 70$ and gradually approaches the performance upper bound  achieved  with  perfect CSI\footnote{ For {\color{black}a fair comparison}, we assume that the training overhead of this upper bound benchmark scheme is the same {\color{black}as for the estimated CSI case.}}. {\color{black}Hence}, Figs. \ref{NMSEvsT} and \ref{TvsRate} both suggest that the MO-EST algorithm can be regarded as a performance benchmark for  {\color{black}sufficiently large} $T$, while the CS-EST algorithm {\color{black}achieves} high performance, especially when the budget for training overhead is limited.
\setlength{\abovecaptionskip}{-1pt}
\begin{table}[]
\caption{ Number of required FLOPs for all simulated algorithms.}\label{Tab:FLOPs2}
\centering
\begin{tabular}{|c|c|c|c|c|}
\hline
Algorithm       & CS-EST        & GAMP        & MO-EST        & PARAFAC        \\ \hline
FLOPs  & $1.50\times 10^8$  & $1.51\times 10^{11}$ &  $8.52\times 10^{8}$  & $3.54\times 10^8$  \\ \hline

\end{tabular}
\end{table}
 \subsection{System Performance Versus  PNR}
\begin{figure}[t]
\centering
\begin{minipage}[t]{0.49\textwidth}
	\centering
 		\includegraphics[width=2.7in]{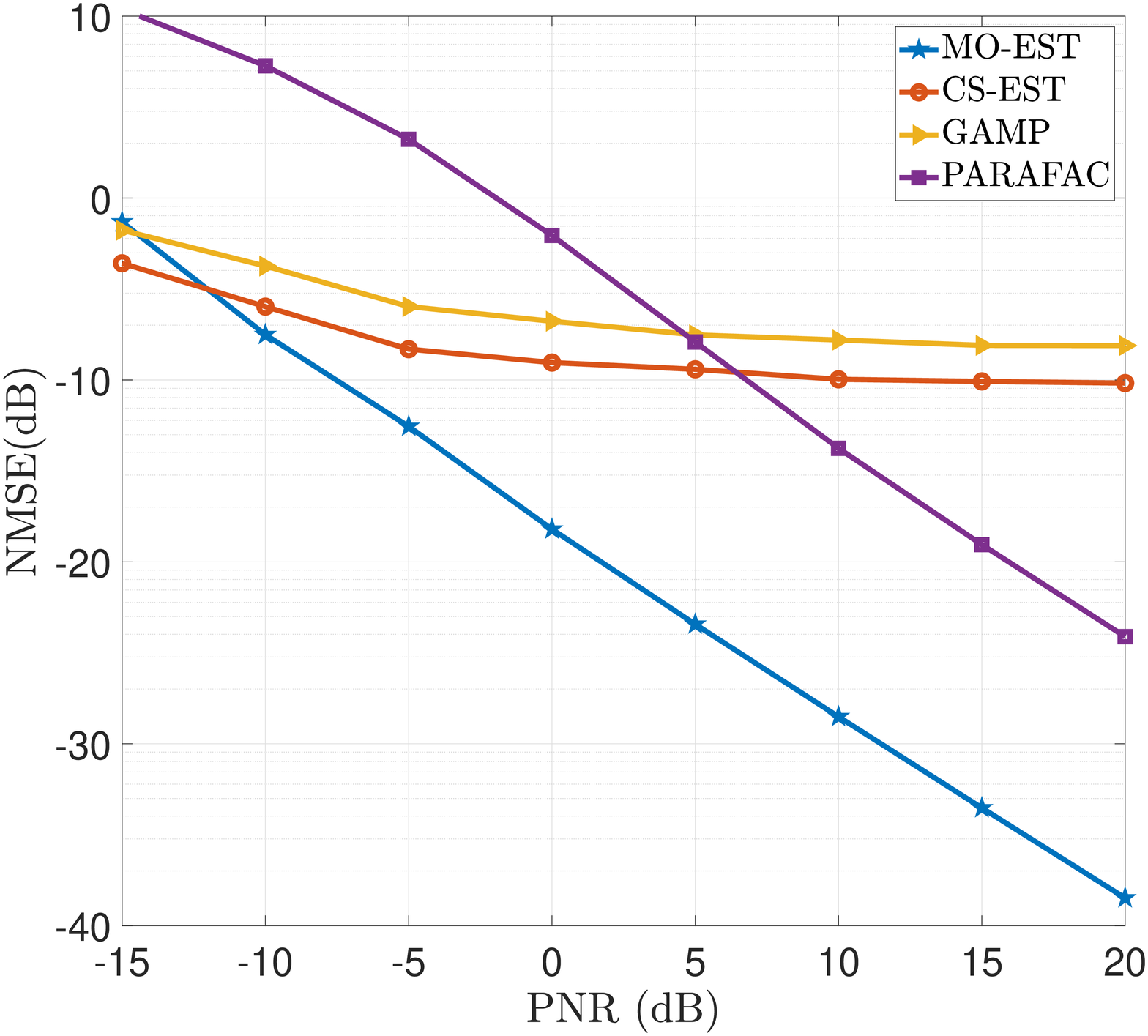}
 	\caption{NMSE versus  PNR for different channel estimation algorithms.}
 	    \label{SNRvsNMSE}
\end{minipage}
\begin{minipage}[t]{0.49\textwidth}
	\centering
 		\includegraphics[width=2.7in]{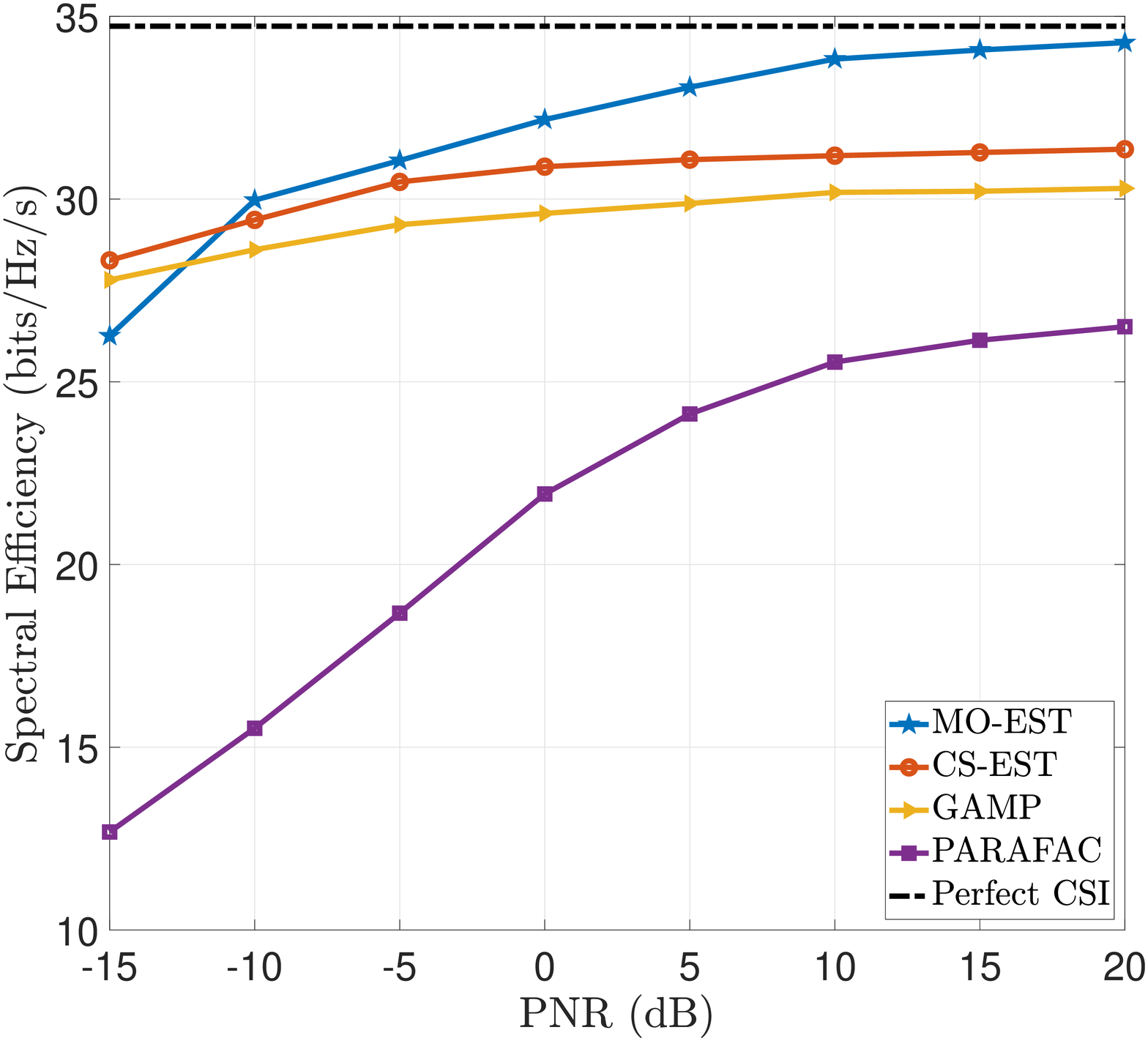}
 	\caption{Spectral efficiency versus PNR for different channel estimation algorithms when $\mathrm{SNR}=10$ dB.}
 	    \label{SNRvsRate}
\end{minipage}
\end{figure}

 Next, in Figs. \ref{SNRvsNMSE} and  \ref{SNRvsRate}, we  show the NMSE  and the  spectral efficiency  versus the PNR, respectively. For the MO-EST,  CS-EST, and  GAMP algorithms, the training overhead $T$ is set to  $200$. On the other hand, we set $T$ to  the minimum required value of $T=MN_t=576$ for the PARAFAC algorithm \cite{araujo_parafac-based_2020}.  Nevertheless, as can be observed, there is a large performance gap between the PARAFAC and the MO-EST {\color{black}algorithms}. This is mainly because  the proposed MO-EST algorithm efficiently exploits the sparsity of mm-wave channels, which highlights the importance of incorporating the rank constraints and $\ell_1$-norm regularizations into the channel estimation algorithm design for IRS-assisted mm-wave MIMO systems. Meanwhile, it can be seen that for the CS-EST  and  GAMP algorithms, the performance gains achieved by increasing the PNR are relatively small. This is  because the main bottleneck in the high PNR regime is the adopted limited angular resolutions. Furthermore, {\color{black}for} the  parameter settings of Figs. \ref{SNRvsNMSE} and  \ref{SNRvsRate}, we list the computational complexity of all investigated algorithms in Table II. The results in Table II, Fig. \ref{SNRvsNMSE}, and  Fig. \ref{SNRvsRate} clearly illustrate the superiority of the proposed MO-EST algorithm especially for channel estimation with high PNRs, which comes at the expense of a higher computational complexity compared to the CS-EST algorithm. On the other hand, the CS-EST algorithm achieves an excellent trade-off between estimation performance and computational complexity.
 \setlength{\abovecaptionskip}{-7pt}

\subsection{Robustness of the Proposed Channel Estimation Algorithms}
\begin{figure}[!t]
 		\centering
 		\includegraphics[width=2.70in]{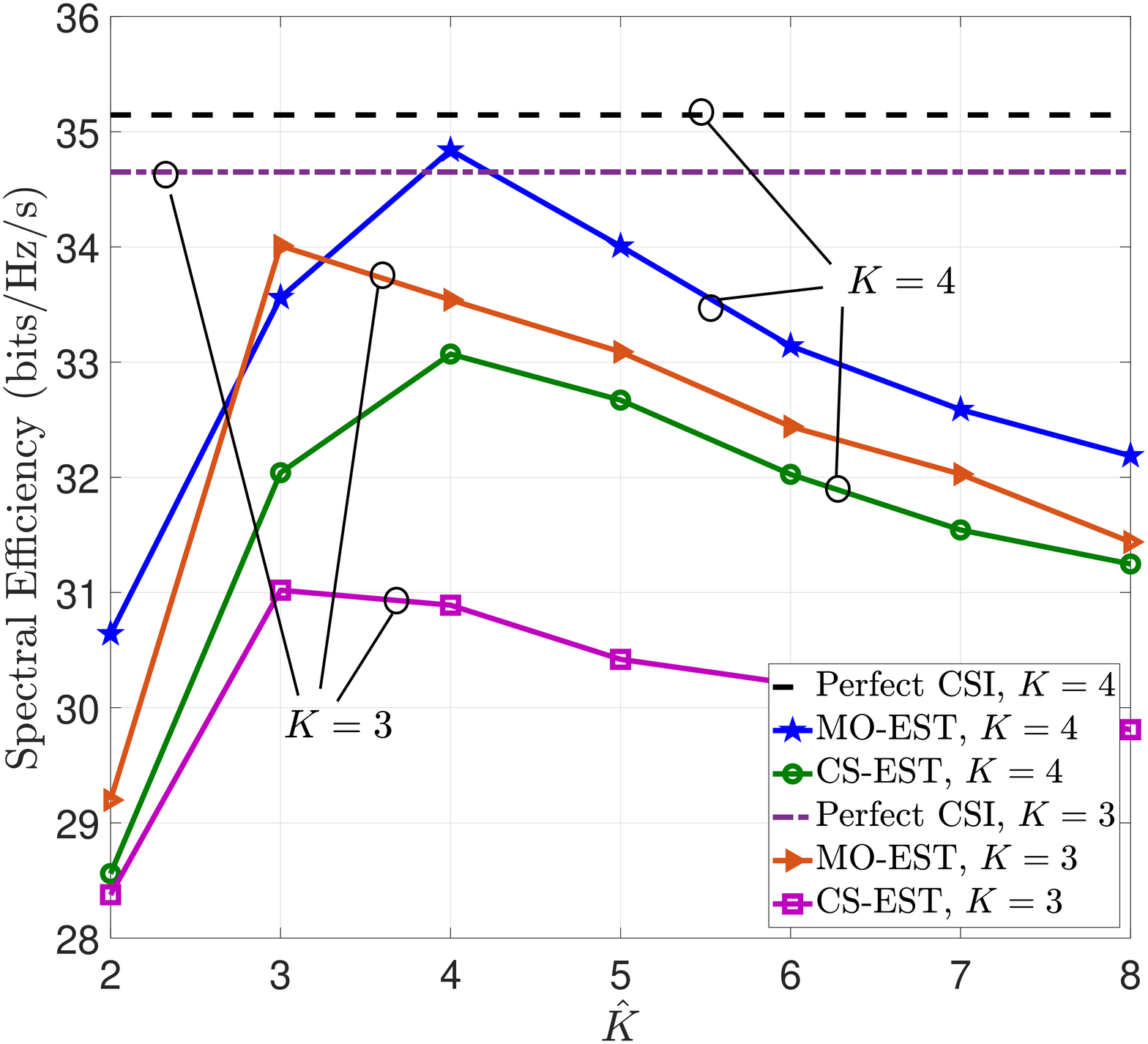}
 	\caption{Spectral efficiency versus $\hat{K}$ for the proposed estimation algorithms when $N_\mathrm{s}=3$.}
 	    \label{Ceps}
 	    \vspace{-10pt}
 	\end{figure}

Finally, in Fig. \ref{Ceps}, we consider the case where the number of paths, $K$, is not perfectly known for channel estimation and test the robustness of the MO-EST {\color{black}and CS-EST} algorithms with respect to the resulting uncertainty, when $\mathrm{PNR}=\mathrm{SNR}=10$ dB and $T=200$. As can be observed,  the proposed MO-EST  and {\color{black} CS-EST} algorithms achieve the highest spectral efficiency  when $\hat{K}=K$, i.e., the number of paths is perfectly known. {\color{black}On the other hand,  a} mismatch between the estimated $\hat{K}$  and the true value of $K$ leads to a performance loss, which, nevertheless,  is limited especially when $\hat{K}\ge K$. 
In particular, for the MO-EST algorithm, the channel matrix ${\mathbf{H}}_\mathrm{c}$ and its estimate $\hat{\mathbf{H}}_\mathrm{c}$ can be decomposed via  SVD, i.e., {\color{black}$\mathbf{H}_\mathrm{c}=\Sigma_{k=1}^{K}\zeta_k\mathbf{u}_k\mathbf{q}_k^H$ and $\hat{\mathbf{H}}_\mathrm{c}=\Sigma_{k=1}^{\hat{K}}\hat{\zeta}_k\hat{\mathbf{u}}_k\hat{\mathbf{q}}_k^H$, where $\zeta_k$ ($\hat{\zeta}_k$), $\mathbf{u}_k$ ($\hat{\mathbf{u}}_k$), and $\mathbf{q}_k$ ($\hat{\mathbf{q}}_k$) denote the ordered singular values, left singular vectors, and right singular vectors, respectively}. In order to minimize  the  objective function in (\ref{eqn:formulate-l1}) based on the LS criterion, the MO-EST algorithm  tries to choose   the ${K}$ largest  singular values of $\hat{\mathbf{H}}_\mathrm{c}$ and the corresponding singular vectors close to the true values while keeping the remaining $\hat{K}-K$ singular values small. In other words, the solution obtained by the MO-EST algorithm satisfies $\hat{\zeta}_k \approx {\zeta}_k$, $\hat{\mathbf{u}}_k \approx \mathbf{u}_k$, $\hat{\mathbf{q}}_k \approx \mathbf{q}_k$ for $k=1,\ldots,K$, and $\hat{\zeta}_k \approx 0$ for $k=K+1, \ldots,\hat{K}$, which leads to a satisfactory estimation performance when $\hat{K}\ge K$. Hence, the proposed MO-EST algorithm is robust with respect to imperfect knowledge of the exact numbers of paths of the estimated channels. Furthermore, for the CS-EST algorithm, since the OMP method itself selects the $K$ columns  most relevant to the residual, the principle components of the channel are not omitted when $\hat{K}\ge K$, and thus, the performance loss is also limited.

\section{Conclusions}
In this paper, we investigated the channel estimation problem for IRS-assisted  mm-wave MIMO systems. By exploiting the sparsity of the mm-wave channel, an  MO-based alternating optimization algorithm and a CS-based algorithm, i.e., the MO-EST algorithm and the CS-EST algorithm, were developed to effectively estimate the IRS-assisted channels. Meanwhile, exploiting the  channel estimates, we also proposed a novel downlink passive beamforming algorithm for maximization of the spectral efficiency by solving an equivalent WMMSE problem. Simulation results showed the   performance improvements achieved with the proposed estimation and  beamforming algorithms compared to several state-of-the-art benchmark schemes. For high PNRs and sufficient number of pilots, the MO-EST algorithm achieves the best performance. On the other hand, the CS-EST algorithm strikes a good balance between the achievable performance, computational complexity, and training overhead. Furthermore, both  proposed algorithms are robust against imperfect knowledge of the sparsity level of the channels. Extending the proposed schemes to multi-user and broadband scenarios {\color{black}are interesting directions} for future research. 
%=======================================
%               Appendix
%=======================================

 \begin{appendices}
      \section{}
We first prove that $\mathrm{rank}(\mathbf{G})=P$. As all AoDs of the IRS-BS channel are different,     matrices $\mathbf{A}_y = \left[\mathbf{a}_y\left(\theta^{1}_\mathrm{t}, \phi^{1}_\mathrm{t}\right),  \ldots, \mathbf{a}_y\left(\theta^{P}_\mathrm{t}, \phi^{P}_\mathrm{t}\right)\right]\in \mathbb{C}^{M_y\times P}$ and  $\mathbf{A}_z = \left[\mathbf{a}_z(\phi^{1}_\mathrm{t}),   \ldots, \mathbf{a}_z(\phi^{P}_\mathrm{t})\right]\in \mathbb{C}^{M_z\times P}$ are both Vandermonde matrices, whose column vectors are linearly independent. Therefore,  matrix $\mathbf{A}_\mathrm{t} = \left[\mathbf{a}_\mathrm{t} \left(\theta^{1}_\mathrm{t}, \phi^{1}_\mathrm{t}\right),  \ldots, \mathbf{a}_\mathrm{t} \left(\theta^{P}_\mathrm{t}, \phi^{P}_\mathrm{t}\right)\right]\in \mathbb{C}^{M\times P}$ with linearly independent columns $\mathbf{a}_\mathrm{t} \left(\theta^{p}_\mathrm{t}, \phi^{p}_\mathrm{t}\right)=\mathbf{a}_y\left(\theta^{p}_\mathrm{t}, \phi^{p}_\mathrm{t}\right)\otimes \mathbf{a}_z(\phi^{p}_\mathrm{t}),$ for $p=1,\dots, P$, satisfies that $\mathrm{rank}(\mathbf{A}_\mathrm{t})=P$. Similarly, $\mathbf{A}_\mathrm{r}=[\mathbf{a}_\mathrm{r}(\theta^{1}_\mathrm{r}),   \ldots,\\ \mathbf{a}_\mathrm{r}(\theta^{P}_\mathrm{r})]\in \mathbb{C}^{N_\mathrm{BS}\times P}$ also  satisfies $\mathrm{rank}(\mathbf{A}_{\mathrm{r}}) = P$.  According to (\ref{channel_model}), $\mathbf{G}$ can be expressed as
 \begin{equation}
 \label{Hr2}
     \mathbf{G} = \mathbf{A}_{\mathrm{r}}\boldsymbol{\Sigma} \mathbf{A}_{\mathrm{t}}^H,
 \end{equation}
 where $\boldsymbol{\Sigma} = \mathrm{diag}(\alpha^1,\ldots, \alpha^P)$ is also a rank-$P$ matrix. According to the   rank properties of matrices \cite{2017Zhangxianda}, we have
 \begin{equation}
 \label{rank-pro}
     \mathrm{rank}(\mathbf{A} \mathbf{B}) \ge \mathrm{rank}(\mathbf{A})+\mathrm{rank}(\mathbf{B})-k, \quad \mathrm{rank}(\mathbf{A} \mathbf{B}) \le \min \{\mathrm{rank}(\mathbf{A}), \mathrm{rank}(\mathbf{B})\},
 \end{equation}
 for arbitrary matrices $\mathbf{A}\in \mathbb{C}^{m\times k}$ and  $\mathbf{B}\in \mathbb{C}^{k\times n}$. Combining the results in (\ref{Hr2}) and (\ref{rank-pro}), we see that  $\mathrm{rank}(\mathbf{G}) = P,$ and similarly, we can prove $\mathrm{rank}(\mathbf{H}) = Q$.
        \section{}
       In order to determine the conjugate gradient $\nabla_{\mathbf{X}_i^*} f_1$, we first compute the differential of $f_1$ with respect to $\mathbf{X}_i^*$. According to some basic differentiation rules for complex-valued matrices \cite{2020ComplexDerive}, we have
       \begin{equation}\label{eqn:app1}
         \mathrm{d}(\|\mathbf{R} - \mathbf{X}\mathbf{F}\|_F^2) = -\mathrm{tr}\left(\mathbf{F}\mathrm{d}(\mathbf{X}^H)\mathbf{R} + \mathbf{F}^H\mathrm{d}(\mathbf{X}^H)\mathbf{X}\mathbf{F}\right).
       \end{equation}
Besides, notice that $\|\boldsymbol{\lambda}_\mathbf{X}\|_1=\sum_i\sum_j{|[\mathbf{A}_\mathrm{BS}^H\mathbf{X}\mathbf{A}_\mathrm{I}]_{ij}|}$, and therefore, defining the $i$-th column of $\mathbf{A}_\mathrm{BS}$ as $\mathbf{a}_i$ and the $j$-th column of $\mathbf{A}_\mathrm{I}$ as $\mathbf{b}_j$, we have
\begin{equation}
\begin{split}\label{eqn:app2}
       &\mathrm{d}(\|\boldsymbol{\lambda}_\mathbf{X}\|_1)\\&=\sum_i\sum_j{\mathrm{d}\left(\left(\mathbf{b}_j^H\mathbf{X}^H\mathbf{a}_i\mathbf{a}_i^H\mathbf{X}\mathbf{b}_j\right)^{\frac{1}{2}}\right)}= \frac{1}{2}\sum_i\sum_j{\left(\mathbf{b}_j^H\mathbf{X}^H\mathbf{a}_i\mathbf{a}_i^H\mathbf{X}\mathbf{b}_j\right)^{-\frac{1}{2}}\mathbf{b}_j^H\mathrm{d}(\mathbf{X}^H)\mathbf{a}_i\mathbf{a}_i^H\mathbf{X}\mathbf{b}_j}\\
       &=\frac{1}{2}\mathrm{tr}\left(\sum_i\sum_j{\mathbf{a}_i\mathbf{a}_i^H\mathbf{X}\mathbf{b}_j\left(\mathbf{b}_j^H\mathbf{X}^H\mathbf{a}_i\mathbf{a}_i^H\mathbf{X}\mathbf{b}_j\right)^{-\frac{1}{2}}\mathbf{b}_j^H\mathrm{d}(\mathbf{X}^H)}\right)
       =\frac{1}{2}\mathrm{tr}\left(\mathbf{A}_\mathrm{BS}\mathbf{Y}\mathbf{A}_\mathrm{I}^H\mathrm{d}(\mathbf{X}^H)\right),
\end{split}
\end{equation}
where $\mathbf{Y}$ is defined in \eqref{eqn:Y}. Then, combining the results in \eqref{eqn:app1} and \eqref{eqn:app2}, the differential of $f_1$ with respect to $\mathbf{X}_i^*$ is given by
\begin{equation}
    \mathrm{d}(f_1) = \mathrm{tr}\left(\left(-\mathbf{R}\mathbf{F}^H + \mathbf{X}\mathbf{F}\mathbf{F}^H +  \frac{\mu_\mathbf{G}}{2}\mathbf{A}_\mathrm{BS}\mathbf{Y}\mathbf{A}_\mathrm{I}^H\right)\mathrm{d}(\mathbf{X}^H)\right).
\end{equation}
Finally, considering the fact that $\mathrm{d}(f_1) = \mathrm{tr}\left(\nabla_{\mathbf{X}_i^*} f_1\mathrm{d}(\mathbf{X}^H)\right)$, the proof is completed.
  \end{appendices}

%=======================================
%               Reference
%=======================================
\newpage
\bibliography{LT}
\bibliographystyle{IEEEtran}	
\end{document}